\documentclass[aos]{imsart}

\RequirePackage{amsthm,amsmath,amsfonts,amssymb}
\RequirePackage[numbers,sort&compress]{natbib}
\RequirePackage[colorlinks,citecolor=blue,urlcolor=blue]{hyperref}
\RequirePackage{graphicx}

\usepackage{subcaption}
\usepackage{tikz}
\usepackage{pgfplots}
\pgfplotsset{compat=1.17}
\usepackage{enumerate}
\usepackage{booktabs} 

\newcommand{\ARL}{\mathrm{ARL}}

\newcommand{\Tcal}{\mathcal T}

\newcommand{\Chan}{\mathrm{Chan}}
\newcommand{\CC}{\mathrm{CC}}
\newcommand{\Sum}{\mathrm{sum}}
\newcommand{\Min}{\mathrm{min}}

\newcommand{\SSBH}{\mathrm{SSBH}}

\DeclareMathOperator*{\argmax}{arg\,max}

\newcommand{\Unif}{\mathrm{Unif}}

\newcommand{\Prp}[1]{\Pr\left[#1 \right]}

\newcommand{\simiid}{\overset{\mathsf{iid}}{\sim}}

\newcommand{\Ncal}{\mathcal{N}}

\newcommand{\HC}{\mathrm{HC}}

\newcommand{\LR}{\mathsf{LR}}
\newcommand{\GLR}{\mathsf{GLR}}

\newcommand{\ex}[1]{\ensuremath{\mathbb{E}\left[ #1\right]}}
\newcommand{\exsub}[2]{\ensuremath{\mathbb{E}_{#1}\left[ #2\right]}}

\newcommand{\sigTwoColor}{blue!60!white}
\newcommand{\sigHalfColor}{green!60!black}
\newcommand{\sigOneColor}{red!90!black}

\newcommand{\abs}[1]{\ensuremath{\left\vert#1\right\vert}}

\startlocaldefs
\theoremstyle{plain}

\newtheorem{theorem}{Theorem}[section]
\newtheorem{corollary}[theorem]{Corollary}
\newtheorem{lemma}[theorem]{Lemma}
\theoremstyle{remark}

\endlocaldefs

\begin{document}

\begin{frontmatter}
\title{Higher-criticism for sparse multi-stream\\ change-point detection}
\runtitle{Higher-criticism for sparse multi-stream change-point detection}

\begin{aug}
\author[A]{\fnms{Tingnan}~\snm{Gong}\ead[label=e1]{tgong33@gatech.edu}\orcid{0000-0002-1542-5787}},
\author[B]{\fnms{Alon}~\snm{Kipnis}\ead[label=e2]{alon.kipnis@runi.ac.il}\orcid{0000-0003-3798-8035}}
\and
\author[A]{\fnms{Yao}~\snm{Xie}\ead[label=e3]{yao.xie@isye.gatech.edu}\orcid{0000-0001-6777-2951}}\footnote{Authors are listed in alphabetical order.}
\address[A]{
H. Milton Stewart School of Industrial and Systems Engineering,\\ 
Georgia Institute of Technology\printead[presep={,\ }]{e1,e3}}

\address[B]{Efi Arazi School of Computer Science, Reichman University\printead[presep={,\ }]{e2}}
\end{aug}

\begin{abstract}
We study a statistical procedure based on higher criticism (HC) to address the sparse multi-stream quickest change-point detection problem. Namely, we aim to detect a potential change in the distribution of multiple data streams at some unknown time. If a change occurs, only a few streams are affected, whereas the identity of the affected streams is unknown. The HC-based procedure involves testing for a change point in individual streams and combining multiple tests using higher criticism. Relying on HC thresholding, the procedure also indicates a set of streams suspected to be affected by the change. We provide a theoretical analysis under a sparse heteroscedastic normal change-point model. We establish an information-theoretic detection delay lower bound when individual tests are based on the likelihood ratio or the generalized likelihood ratio statistics and show that the delay of the HC-based method converges in distribution to this bound. In the special case of constant variance, our bound coincides with known results in (Chan, 2017) \cite{chan2017optimal}. We demonstrate the effectiveness of the HC-based method compared to other methods in detecting sparse changes through extensive numerical evaluations. 
\end{abstract}

\begin{keyword}[class=MSC]
\kwd{62L10}
\end{keyword}

\begin{keyword}
\kwd{Higher-criticism}
\kwd{change-point detection}
\kwd{multi-stream}
\end{keyword}

\end{frontmatter}

\section{Introduction}

Sequential change-point detection, which aims to identify changes in the distribution of streaming data as quickly as possible, is a fundamental challenge in statistics \cite{xie2021sequential}. Recently, there has been much interest in the settings where a large number of data streams are being monitored and the potential change may affect only a small subset of variables, a scenario referred to as {\it sparse} multi-stream change-point detection \cite{tartakovsky2008asymptotically,xie2013sequential,chan2017optimal,gosmann2022sequential,hu2023likelihood}. The challenges in this situation follow from the unknown identity of the affected streams if they exist, their relatively small number, and the standard sequential change-point tradeoffs of minimizing {\it detection delay} while maintaining the longest possible run without false positives under the null. 

In offline hypothesis testing, Higher Criticism (HC) \cite{donoho2004higher} has proven to be an effective tool for global testing against sparse alternatives 
\cite{donoho2009feature,donoho2015higher,li2015higher,DonohoKipnis2020}. The relative simplicity of HC, its well-understood properties, and its inherent feature selection mechanism \cite{donoho2008higher} have made HC a popular choice for sparse signal detection among practitioners
\cite{donoho2015higher,kipnis2022higher}. 
Therefore, we aim to develop an adaptation of HC for sparse sequential change-point detection. 

In this paper, we present
a new procedure for detecting a change point affecting a small and unknown subset of data streams by combining their detection statistics using HC. 
Our procedure can be viewed as a sequential adaptation of offline sparse signal detection using HC: in every time instance, we test the significance of a change in every stream individually using one of the known tests, such as the cumulative sum (CUSUM) test, then use HC to combine P-values from many individual change-point tests. The resulting statistic is sensitive to sparse changes. Namely, changes affecting only a small fraction of the streams when the locations of the affected streams are unknown. Additionally, HC provides a mechanism to identify streams suspected to be affected, offering optimality properties compared to other fault localization methods based on feature selection methods such as those based on false discovery rate controlling \cite{donoho2008higher,donoho2009feature}. Since the procedure wraps around P-values, it can be naturally applied to data streams of heterogeneous pre- and post-change distributions in individual streams with general change-point detection statistics. We demonstrate the effectiveness of the HC-based procedure through extensive numerical evaluations and provide a theoretical analysis under a sparse multi-stream model with normal data, where individual streams may experience changes in their means and variances, referred to as the heteroscedastic change. We show that the expected detection delay of HC of P-values obtained from the likelihood ratio (LR) based CUSUM \cite{lorden1971procedures}, or from the generalized likelihood ratio (GLR) test \cite{siegmund1995using}, attains the information-theoretic limits of detection when only the mean is affected (i.e., the homoscedastic change) derived in \cite{chan2017optimal}. Our results show that the HC of P-values from individual streams is also optimally adaptive to the sparsity level. Our key contributions are summarized as follows. 
\begin{itemize}
\item We propose a \emph{sequential} change-point detection method based on HC, effectively detecting a global change point with sensitivity to a change affecting a few streams simultaneously out of many monitored streams. 
\item We consider a multi-stream normal sparse heteroscedastic change point model and characterize the information-theoretic detection delay in this model for individual tests based on the LR and the GLR statistics. We deliver such results concerning the detection delay rather than the expected detection delay studied in \cite{chan2017optimal}. 
\item We show that the HC-based procedure attains the asymptotic minimal detection delay in the model mentioned above, even when the sparsity and post-change mean and variance are unknown.
\item We use extensive simulations to compare the detection procedure based on HC to other procedures proposed in the literature for the same setting in finite sample sizes. Our evaluations show that the HC-based procedure typically outperforms other methods. 
\end{itemize}
Our procedure may also benefit from other well-known properties of HC, such as robustness to mild dependency among the P-values
\cite{delaigle2011robustness,hall2008properties} and a known procedure for adapting to stronger dependencies \cite{hall2010innovated}.

\vspace{.1in}

To the best of our knowledge, this paper is the first to study the use of higher criticism statistics for combining P-values to detect sparse changes sequentially.
Related recent works focusing on offline change point and anomaly detection in multi-stream data include \cite{chan2015optimal,bhamidi2018change,pilliat2023optimal,stoepker2024anomaly}. Methods for sparse sequential change-point detection include tests based on the stream providing the strongest indication \cite{tartakovsky2006detection,gosmann2022sequential}, combining local tests (typically, CUSUM-based) via averaging or other methods \cite{mei2010efficient,tartakovsky2008asymptotically,liu2019scalable,xu2022active}, kernel-based scan statistics \cite{liu2019scalable}, and a mixture of likelihood ratios known to be useful when the number of affected streams is known 
\cite{xie2013sequential}. Methods with proven optimal adaptivity to the sparsity level include the mixture likelihood ratio with adaptive window size of \cite{chan2017optimal}, and 
combining P-values from individual streams using likelihood scores \cite{hu2023likelihood}.  

The rest of the paper is organized as follows. Section~\ref{sec:setup} sets up the data analysis problem. Section~\ref{sec:global_detection} presents the HC procedure for sequential change-point detection. Section~\ref{sec:theory} contains the main theoretical results. Section~\ref{sec:experimnets} contains numerical examples that validate the theoretical results and compare HC's performance with other methods. Section~\ref{sec:discussion} summarizes the paper. Section~\ref{sec:proof} contains proofs. Some proof details are delegated to the Appendix.

\section{Method}
\label{sec:global_detection}

\subsection{Setup}
\label{sec:setup}
We observe $N$ data streams over time. 
Let $X_{n,t}$ represent the observation of the $n$-th stream at time $t=1,2,\ldots$. The null hypothesis states no change in the distribution of any stream. Under the alternative hypothesis, there exists an \emph{unknown} time $\tau$, $0 < \tau <\infty$, and an \emph{unknown} subset of streams with index set $I \subset \{1,\ldots,N\}$ affected by the change at time $t=\tau$ and thus their distribution is different at times $t=\tau,\tau+1,\ldots$. In contrast, the distribution of each unaffected stream remains unchanged. Formally, given sequences of pre-change distributions $\{F^0_{n}\}_{n=1}^N$ and post-change distributions $\{F^1_{n}\}_{n=1}^N$, we consider the following sequential hypothesis testing problem:
\begin{align}
\begin{split}
\label{eq:hyp_test}
    H_0: &\quad X_{n,t} \sim F^0_n,\quad  t = 1, 2, \ldots \\
    H_1: & \quad X_{n,t} \sim  \begin{cases}
    F^0_{n}, & n \notin I, \\
    F^0_{n}, & n \in I,\quad t = 1, \ldots, \tau-1, \\
    F^1_{n}, & n \in I,\quad t = \tau, \tau+1, \ldots.
    \end{cases}
\end{split}
\end{align}
Our goal is to detect the change as quickly as possible after it occurs while allowing a sufficiently long run without a false alarm when there is no change. We focus on the case when the fraction of affected streams $p:=|I|/N$ is small. Our approach involves testing for a change in every stream individually and combining multiple tests using {\it higher-criticism} \cite{donoho2004higher,donoho2015higher}.

\subsection{Change-point testing in individual streams}

Given the observations at time $t$, suppose we use certain detection statistics $Y_{n,t}$ that is a function of $\{X_{n,s}\}_{s \leq t}$ for sequential change-point detection at the $n$-th stream. We summarize the significance of the change by a P-value with respect to the no-change distribution $F_n^0$. 
Let $\{\pi_{n,t}\}$, $n=1,\ldots,N$ be a sequence of such P-values at times $t=1, 2\ldots$.

\subsection{Detection using Higher Criticism}

\begin{figure}[b!]
\centering 
\begin{subfigure}[h]{0.45\linewidth}
\includegraphics[width=\linewidth]{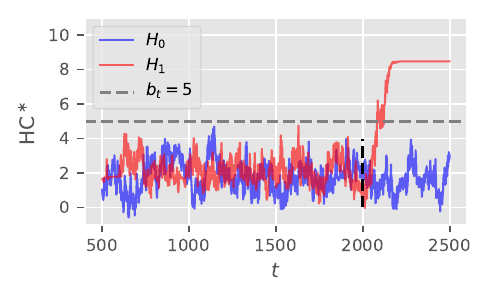}
\caption{}
\label{fig:exact-CUSUM-HC-r0.05-beta0.5}
\end{subfigure}
\begin{subfigure}[h]{0.45\linewidth}
\includegraphics[width=\linewidth]{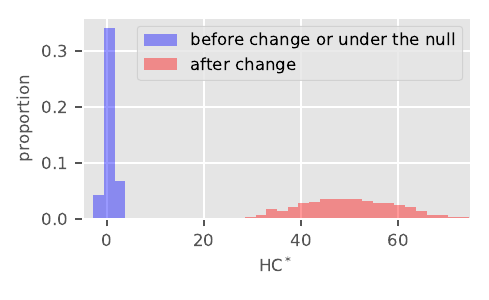}
\caption{}
\label{fig:exact-CUSUM-HC-r0.05-beta0.5-hist}
\end{subfigure}
\caption{HC detection statistic $\HC^*_t$ of \eqref{eq:HC_t_star} before and after the change that occurs at time $t=2000$. Each of $N=500$ streams is distributed as $\Ncal(0,1)$ before the change, after the change, streams in $I$ are distributed as $\Ncal(\mu_r(N),1)$. P-values of each stream at each time are computed using the CUSUM statistic \eqref{eq:YLR_def} under the null distribution (before the change). (a) Trajectories of $\HC^*_t$ over $t$, when there is no change, and when there is a change at 2000, post-change distribution being normally distributed with $\mu_r(N) = 0.35$ and $|I|=4$. (b) Histograms of 1000 simulated values of $\HC_t^*$ at time $t=1,000$ (before the change) and $t=2,100$ (after the change), with $\mu_r(N) = 0.79$ and $|I|=22$. \label{fig:trajectory_example}
}
\end{figure}

Before presenting the change point detection procedure, let us review the HC statistic \cite{donoho2004higher,donoho2015higher}. For a given collection of P-values $\pi_1,\ldots,\pi_N$, let $\pi_{(1)}\le \cdots\le \pi_{(N)}$ be their ordered version. The HC statistic of a collection of P-values is defined as
\begin{equation}\label{eq:HC_star}
    \HC^*(\pi_1,\ldots,\pi_N) := 
    \max_{1 \leq n \leq \alpha_0 N }  
    \sqrt{N} \frac{\frac{n}{N}-{\pi}_{(n)}}
    {\sqrt{\frac nN\left(1-\frac n N\right)}},
\end{equation}
where $\alpha_0 \in (0, 1)$ is a tunable parameter known to have little effect on the large sample behavior of $\HC^{*}$ \cite{donoho2015higher}. 

To apply HC in the sequential setting, we form the detection statistic at time $t$ using individual P-values at each data stream:
\begin{align}\label{eq:HC_t_star}
    \HC^{*}_t = \HC^{*}(\pi_{1,t},\ldots,\pi_{N,t}),\qquad t=1,2\ldots.
\end{align}
We declare global change when the sequence $\{\HC_t^{*}\}_{t=1,2,...}$ exceeds a prescribed threshold sequence $\{b_t\}_{t=1,2,...}$ for the first time. This detection procedure corresponds to the following stopping rule called the {\it HC procedure}:
\begin{align}
\label{eq:HC_stopping_rule}
    T_{\HC} = \inf\{t>0: \HC_t^{*}>b_t\}, 
\end{align}
where $b_t >0$ is a user-specified (usually time-invariant) threshold chosen to control the false alarm when there is no change, and thus control the run length (RL) and the average run length (ARL). Figure~\ref{fig:exact-CUSUM-HC-r0.05-beta0.5} presents an example of the detection statistic and its empirical distributions before and after the change.

The HC procedure can perform change localization conveniently, once a change has been detected. When the detection statistic has exceeded the threshold, thus, a global change is declared, we may select streams suspected to experience a change, by setting
\begin{align}
    n^* = \argmax_{1 \leq n \leq \alpha_0 N} \frac{\frac{n}{N}-\pi_{(n)}}{\sqrt{\frac{n}{N}\left(1-\frac{n}{N}\right)}},
    \label{eq:HC_threshold}
\end{align}
and select the $n$-th stream if  $\pi_i \leq \pi_{(n^*)}$. Prior works \cite{donoho2008higher,donoho2009feature,donoho2015higher,jin2016rare} discuss this selection mechanism and its optimality properties in the context of feature selection for classification.

\section{Asymptotic detection performance in normal data}
\label{sec:theory}

In this section, we analyze the asymptotic performance of the HC procedure when detecting a sparse change in the mean across independent normal data streams. The asymptotic setting assumes that the number of streams $N$ goes to infinity while the other problem parameters, such as the average proportion of affected streams and the magnitude of an individual mean change, are calibrated to $N$. 

Our calibration follows that of (Chan, 2017) \cite{chan2017optimal}, which has been shown to yield a challenging and interesting statistical setup: There exists an ``information-theoretic'' lower bound in the sense that any detection procedure with a controllable false alarm rate has an expected delay exceeding this lower bound in the sparse change setting. The main results of this section state that the higher criticism procedure attains this information-theoretic bound. Additionally, we extend the setting of \cite{chan2017optimal} to include potential change in the variance of affected streams, detection based on the generalized likelihood ratio statistics \cite{siegmund1995using}, and convergence in distribution results rather than convergence of the means as studied in \cite{chan2017optimal}. 

\subsection{Problem formulation}

While our procedure works for general distributions, for theoretical analysis, we make some additional distributional assumptions in addition to \eqref{eq:hyp_test}.
Let $I \subset \{1,\ldots,N\}$ be a random set of indices such that $\Prp{ n \in I} = p$ independently across $n=1,\ldots,N$. The observations obey $X_{n,t}\sim \mathcal N(0,1)$ for all $t$ and $n\notin I$, and for $n\in I$, 
\begin{subequations}
    \label{eq:normal_problem}
\begin{equation}
\label{eq:data_model}
    X_{n,t} \sim 
    \begin{cases}
    \Ncal(0,1), & t=1 \ldots, \tau-1,\\ 
    \Ncal(\mu,\sigma^2), & t=\tau, \ldots,\Tcal,
    \end{cases}
\end{equation}
where the paths $\{X_{n,t}\}_{t\geq 1}$ are independent for different $n=1,\ldots,N$, and $\Tcal$ denotes the time horizon. The global change point hypothesis testing can be cast as:
\begin{equation}
\label{eq:global_model}
\begin{aligned}
& H_0^{(N)}:
& & \tau > \Tcal,  \\
& H_1^{(N)}:
& & \tau \leq \Tcal,
\end{aligned}
\end{equation}
\end{subequations}
i.e. $\tau > \Tcal$ indicates no change. Here, for simplicity, we assume that in \eqref{eq:data_model}, $\mu > 0$ is the common shift in the means of all affected streams and $\sigma > 0$ is associated with a common change in their variances. The homoscedastic case ($\sigma = 1$) corresponds to the setting of \cite{chan2015optimal,chan2017optimal,hu2023likelihood}. 

\subsection{Testing individual streams}
We consider two useful cases of test statistics in individual streams. 

\vspace{3pt}
\noindent{\it Case I: CUSUM for mean-shift} (see, e.g., (Lorden 1971) \cite{lorden1971procedures}). Consider the CUSUM statistic formed on each stream assuming the known post-change mean:
    \begin{align}
    Y_{n,t}^{\LR} &:= 
    \max_{k \leq t}
     V_{n,t,k}, \qquad V_{n,t,k} := 
     \left(S_{n,t} - S_{n,k} - \frac{\mu}{2}(t-k)\right)\mu,
     \label{eq:YLR_def}
    \end{align}
    where $S_{n,t}:= \sum_{j=1}^t X_{n,j}$, 
    and $\mu$ is the assumed post-change mean.
    Note that $Y_{n,t}^{\LR}$ is associated with the likelihood ratio statistic for change in the mean of a normal model from zero to $\mu_N(r)$ while the variance is unchanged. This statistic arises in testing for a mean shift by $\mu$ under a local asymptotically normal family; see the discussions in \cite{tartakovsky2005general,xie2021sequential} and references therein.
    The CUSUM $Y_{n,t}^{\LR}$ can be computed recursively in time $t$, facilitating its online implementation and thus a popular choice for online change-point detection. A P-value $\pi_{n,t}^{\LR}(x)$ for the $n$-th stream corresponds to the survival function of $Y^{\LR}_{n,t}$, as in
\begin{align}
    \label{eq:p-val_def_LR}
\pi_{n,t}^{\LR}(x) & := \Prp{Y_{n,t}^{\LR} \geq x \mid H_0}.
\end{align}

\vspace{3pt}
\noindent{\it Case II: Generalized Likelihood Ratio (GLR).} The GLR statistic to detect some change in the mean (see, e.g., (Siegmund and Venkatraman 1995) \cite{siegmund1995using}) is defined by
\begin{align}
    \label{eq:YGLR_def}
Y_{n,t}^{\GLR} := \max_{t-w<k \leq t} W_{n,t,k},\qquad W_{n,t,k} := 
\frac{\left|S_{n,t} - S_{n,k}\right|}{\sqrt{t-k}}.
\end{align}
where $w>0$ is the window-length; this corresponds to the window-limited GLR \cite{lai1998information}. The GLR statistic cannot be computed recursively, unlike the CUSUM statistic in \eqref{eq:YLR_def}. On the other hand, GLR does not require assuming the post-change parameters,  which are typically hard to specify a priori, and the true change magnitude can change from instance to instance. The window length is typically chosen to be large enough so that it does not affect the detection delay.
This statistic is particularly useful when the post-change mean is unknown. The P-value at time $t$ is given by $Y^{\GLR}_{n,t}$ by
\begin{align}
    \label{eq:p-val_def_GLR}
\pi_{n,t}^{\GLR}(x) & := \Prp{Y_{n,t}^{\GLR} \geq x \mid H_0}.
\end{align}

We have a few remarks:
\begin{itemize}
    \item Statistics based on the cumulative sum $S_{n,t}$ such as $Y_{n,t}^\LR$ and $Y_{n,t}^\GLR$ are widely used in the literature for detecting a change in the mean. They are generally less effective for detecting a particularly small or large change in the variance, hence it is useful to think about $\sigma^2$ in our model \eqref{eq:data_model} as a nuisance parameter. As it turns out, larger values of $\sigma^2$ are asymptotically \emph{favorable} for detecting using $Y_{n,t}^\LR$ and $Y_{n,t}^\GLR$.
    \item It is also possible and sometimes helpful to define the test statistics $V_{n,t}$ and $W_{n,t}$ by replacing $S_{n,t}-S_{n,k}$ with $\max\{S_{n,t}-S_{n,k},0\}$. Such replacement does not appear to change the main conclusions concerning asymptotic detection delays.
    \item In either case $\square \in \{\LR, \GLR\}$ and under either hypothesis, $Y_{n,t}^\square$ and thus $\pi_t^\square(Y_{n,t})$ have continuous distributions, hence $\pi_{n,t}^\square(Y_{n,t})$ is uniformly distributed over $(0,1)$ under $H_0$. 
\end{itemize}

\subsection{Asymptotic detection delay}

We study the detection procedure \eqref{eq:HC_stopping_rule} under an asymptotic setting of $N \to \infty$, while $p$ and $\mu$ are calibrated to $N$. Specifically, we calibrate $p$ as
\begin{subequations}
\label{eq:calibration}
\begin{align}
\label{eq:beta_cond}
 p & = N^{-\beta},
\end{align}
where $\beta \in (0,1)$ is a parameter controlling sparsity, and
\begin{align}\label{eq:mu_r}
 \mu := \mu_r(N) := \sqrt{2 r \log(N)},
\end{align}
$r > 0$ is a fixed parameter controlling the magnitude of the mean shift in affected streams. 
\end{subequations}
Under this calibration, \eqref{eq:data_model} defines a sequence of hypothesis testing problems indexed by $N$. 

We characterize the asymptotic detection delay of procedures based on P-values obtained either from $Y_t^{\LR}$ or $Y_t^{\GLR}$. Define the function
\begin{align}
\label{eq:delta_star_def}
\Delta^\ast(r,\beta,\sigma) := \lceil \rho^*(\beta,\sigma)/r \rceil,
\end{align}
(the smallest integer larger than $\rho^*(\beta,\sigma)/r$) where 
\begin{align*}
\rho^*(\beta,\sigma) & :=
\begin{cases}
     (2 - \sigma^2)(\beta - 1/2), & \frac{1}{2} < \beta < 1- \frac{\sigma^2}{4}, \quad 0<\sigma^2<2,\\
    \left(1-\sigma\sqrt{1 -\beta }\right)^2, &  1- \frac{\sigma^2}{4} \leq  \beta < 1, \quad 0<\sigma^2<2,\\
     0, & \frac{1}{2} < \beta <  1-\frac{1}{\sigma^2}, \quad \sigma^2 \geq 2,\\
    \left(1-\sigma\sqrt{1 -\beta }\right)^2, &  1-\frac{1}{\sigma^2} \leq  \beta < 1, \quad \sigma^2 \geq 2.
    \end{cases}
\end{align*}
The function $\rho^*(\beta,\sigma)$ was first introduced in \cite{tony2011optimal} for describing the detection boundary in heteroscedastic sparse signal detection. Here, we will show that $\Delta^*(r,\beta,\sigma)$ characterize the asymptotic detection delay; Figure~\ref{fig:rho_star} illustrates $\Delta^\ast(r,\beta,\sigma)$ for several varying parameter values. 

\begin{figure}
    \centering

    \begin{subfigure}[h]{0.33\linewidth}
    \centering
    \begin{tikzpicture}
    \begin{axis}[
    width=7cm,
    height=4cm,
    legend style={at={(0,1)},
      anchor=north west, legend columns=1},
    ylabel={\small $\Delta^*(0.05,\beta,\sigma)$},
    xlabel={$\beta$},
    ytick={1,10,20},
    yticklabels={1,,$1/r$},
    xtick={0.45,0.5,0.6,0.7,0.8,0.9,1},
    xticklabels={,0.5,0.6,0.7,0.8, 0.9,1},
    legend cell align={left},
    ymin=1,
    xmin=0.47,
    xmax=1,
    ymax=20,
    ]

\addplot[domain=0.4:0.5, color=\sigHalfColor, style=thick, samples = 13, mark=+, mark size=1pt] {0};
\addlegendentry{\scriptsize $\sigma=1/2$};

\addplot[domain=0.4:0.5, color=\sigOneColor, style=thick,mark=*, samples = 13,  mark size=0.5pt] {0};
\addlegendentry{\scriptsize $\sigma=1$};

\addplot[domain=0.4:0.5, color=\sigTwoColor, style=thick, samples = 13, mark=triangle, mark size=1pt] {0};
\addlegendentry{\scriptsize $\sigma=2$};

\addplot[domain=0.4:0.5, color=\sigOneColor, style=thick, samples = 13] {0};

\def\sig{.5};
\def\rr{0.05}

\addplot[domain=0.5:1-\sig^2/4, color=\sigHalfColor, style= thick, samples = 43,mark=+, mark size=1pt] {round((2-\sig^2)*(x-1/2)/\rr + 0.5)};

\addplot[domain=1- \sig^2/4:1, samples = 11, color=\sigHalfColor, style= thick,mark=+, mark size=1pt]
    {round((1-\sig*(1-x)^0.5)^2/\rr+0.5)};


\def\sig{1};
\def\rr{0.05}

\addplot[domain=0.5:1-(\sig^2)/4, color=\sigOneColor, style= thick, samples = 23, mark=*, mark size=0.5pt] {round((2-\sig^2)*(x-1/2)/\rr + 0.5)};

\addplot[domain=1- \sig^2/4:1, samples = 31, color=\sigOneColor, style= thick, mark=*, mark size=0.5pt]
    {round((1-\sig*(1-x)^0.5)^2/\rr+0.5)};

\def\sig{2};
\def\rr{0.05}

\addplot[domain=0.5:1-1/\sig^2, color=\sigTwoColor, style= thick, samples = 20, mark=triangle, mark size=1pt] {round(1)};

\addplot[domain=1- 1/\sig^2:1, samples = 31, color=\sigTwoColor, style= thick, mark=triangle, mark size=1pt]
    {round((1-\sig*(1-x)^0.5)^2/\rr+0.5)};



\node[left] (topl) at (axis cs:0.6,1.1) {};
\node[right] (botr) at (axis cs:0.94,0.05) {};

\end{axis}

\end{tikzpicture}
    \caption{}
    \end{subfigure}
    \hspace{+50pt}
    \begin{subfigure}[h]{0.33\linewidth}
    \centering
    \begin{tikzpicture}
    \begin{axis}[
    width=7cm,
    height=4cm,
    legend style={at={(1,1)},
      anchor=north east, legend columns=1},
    ylabel={\small $\Delta^*(r,\beta,1)$},
    xlabel={$r$},
    ytick={0,5,10,15,20,25},
    yticklabels={,5,10,,20,},
    xtick={0.05,0.1,0.2,0.3},
    xticklabels={,0.1,0.2,0.3},
    legend cell align={left},
    ymin=0,
    xmin=0,
    xmax=0.25,
    ymax=30,
    ]

\def\sig{1};
\def\be{0.6}
\addplot[domain=0.001:0.3, color=\sigTwoColor, style= thick, samples = 57, mark=triangle, mark size=1pt] {round((2-\sig^2)*(\be-1/2)/x + 0.5)};
\addlegendentry{\scriptsize $\beta=0.6$}

\def\be{0.75}
\addplot[domain=0.001:0.3, color=\sigOneColor, style= thick, samples = 57, mark=*, mark size=1pt] {round((2-\sig^2)*(\be-1/2)/x + 0.5) };
\addlegendentry{\scriptsize $\beta=0.75$}

\def\be{0.9}
\addplot[domain=0.001:0.3, samples = 27, color=\sigHalfColor, style= thick, mark=square, mark size=1pt] {round((1-\sig*(1-\be)^0.5)^2/x+0.5)};
\addlegendentry{\scriptsize $\beta=0.9$}

\end{axis}

\end{tikzpicture}
    \caption{}
    \end{subfigure}
    
    \caption{Curves describing the asymptotic theoretical detection delay $\Delta^*(r,\beta,\sigma)$ of \eqref{eq:delta_star_def} in multistream normal data with a heteroscedastic sparse change for detection based on LR or GLR statistics (Theorems \ref{thm:main_impossible} and \ref{thm:HC}).
    (a): $\Delta^*(r,\beta,\sigma)$ versus $\beta$ for several values of $\sigma$. 
    (b): $\Delta^*(r,\beta,\sigma)$ versus $r$ for several values of $\beta$. The detection delay increases with larger $\beta$ (higher sparsity) and decreases with larger $r$ (stronger change magnitude) and larger $\sigma^2$ (larger post-change variance).
    }
    \label{fig:rho_star}
\end{figure}

Our first main result says that for detection procedures based on LR or GLR tests, the probability of detecting at times $t < \tau+\Delta^*(r,\beta,\sigma)$ is asymptotically identical under the null and the alternative. This result implies that the detection delay allowing for a meaningful change-point monitoring procedure is bounded below by $\Delta^*(r,\beta,\sigma)$.
\begin{theorem}
\label{thm:main_impossible}
Consider the change-point detection problem \eqref{eq:normal_problem} and P-values
\[
\pi_{n,t} := \pi^{\square}_t( Y^{\square}_{n,t}), \quad n = 1,\ldots,N, \quad t = 1,\ldots,\Tcal,
\]
where $\square \in \{\LR, \GLR\}$ as defined by  
\eqref{eq:YLR_def}-\eqref{eq:p-val_def_LR} or \eqref{eq:YGLR_def}-\eqref{eq:p-val_def_GLR}. For a given test statistic $U_t$ based on $\pi_{1,t},\ldots,\pi_{N,t}$, consider a detection procedure that stops at time $T_U$ as soon as $U_t$ exceeds $b_t^{(N)}$. Consider an asymptotic setting $N \to \infty$, with $p$ and $\mu$ calibrated to $N$ as in \eqref{eq:calibration} and $\Tcal/\log(N) \to 0$. If $t < \tau+\Delta^*(r,\beta,\sigma)$, for any array of thresholds $\{b_t^{(N)},\,t=1,\ldots,\Tcal,\,N=1,2,\ldots\}$ we have
\[
\Prp{ T_U \leq t \mid H_1^{(N)} } - \Prp{ T_U \leq t \mid H_0^{(N)} } \to 0.
\]
\end{theorem}
In other words, Theorem~\ref{thm:main_impossible} says that whenever there exists a procedure for detecting a change 
based on the LR or GLR statistics with a probability of true detection at time $t$ asymptotically exceeding the probability of a false alarm at time $t$, then $t\geq \Delta^*(r,\beta,\sigma)$.

The following result shows that it is possible to detect using higher criticism with a delay converging to $\Delta^*(r,\beta,\sigma)$. 
\begin{theorem}
\label{thm:HC}
Consider the change-point detection problem \eqref{eq:data_model} and P-values
\[
\pi_{n,t} := \pi^{\square}_t( Y^{\square}_{n,t}), \quad n = 1,\ldots,N, \quad t = 1,\ldots,\Tcal,
\]
where $\square \in \{\LR, \GLR\}$ as defined by  
\eqref{eq:YLR_def}-\eqref{eq:p-val_def_LR} or \eqref{eq:YGLR_def}-\eqref{eq:p-val_def_GLR}. Let the detection procedure stop at time $T_{\HC}$ as soon as $\HC_t^* = \HC^*( \{\pi_{n,t}^{\square}\}_{n=1}^N)$ exceeds $b_t$:
\[
    T_{\HC} = \inf\{t: \HC_t^* > b_t\}.
\]
In an asymptotic setting $N \to \infty$, with $p$ and $\mu$ calibrated to $N$ as in \eqref{eq:calibration} and $\Tcal/\log(N) \to 0$, for any integer $\Delta \geq \Delta^*(r,\beta,\sigma)$, there exists an array of thresholds $\{b_t^{(N)},\,t=1,\ldots,\Tcal,\, N=1,2,\ldots\}$ such that, 
\[
\Prp{ T_{\HC}-\tau = \Delta \mid H_1^{(N)} } \to 1,
\]
while 
\[
\Prp{ T_{\HC} \leq \Tcal \mid H_0^{(N)}} \to 0.
\]
\end{theorem}
To state the following corollary, we first define a  false alarm rate $\alpha$ for a finite time horizon $\mathcal T$, as $\Prp{ T_{\HC} \leq \Tcal \mid H_0^{(N)}}\leq \alpha$.
\begin{corollary}
\label{cor:HC}
    Under the setting and conditions of Theorem~\ref{thm:HC}, for any $(\beta,\sigma,r)$ configuration such that $\rho^*(\beta,\sigma)/r$ is not an integer, and for any false alarm rate $\alpha \in (0,1)$ in the horizon $\Tcal$, 
    there exists an array of threshold $b_t^{(N)}$, which controls the false alarm at level $\alpha$, 
    such that the detection delay $T_{\HC} - \tau$ converges to $\Delta^*(r,\beta,\sigma)$ in distribution uniformly in $\tau$. In particular,
    \begin{align}
        \max_{1 \leq \tau \leq \Tcal} \exsub{\tau}{ T_{\HC} - \tau \mid T_{\HC} \geq \tau} = \Delta^*(r,\beta,\sigma).
    \end{align}
\end{corollary}

\subsection{Discussions}

\subsubsection{Information-theoretic delay and phase transition}\label{sec:phase}

Theorem~\ref{thm:main_impossible} establishes a fundamental limit on the ability to construct a meaningful detection procedure for delays $\Delta < \Delta^*(r,\beta,\sigma)$, since the probability of a false alarm before such $\Delta$ converges to the probability of true detection before $\Delta$. This renders the detection delay comparable in magnitude to the in-control run length, effectively reducing detection to a random guess and, therefore, making it uninformative. On the other hand, Theorem~\ref{thm:HC} shows that there exist detection procedures with detection delays converging in distribution (from above) to $\Delta^*(r,\beta,\sigma)$ while maintaining a stretch of no false alarms $\Tcal$ going to infinity. One such procedure is the higher criticism-based procedure of \eqref{eq:HC_stopping_rule}. This describes a phase transition in the ability to provide meaningful change-point monitoring: asymptotically impossible for delays smaller than $\Delta^*(r,\beta,\sigma)$ and asymptotically possible for larger delays. Previous results addressing the information-theoretic detection delay focused on the \emph{expected} detection delay (EDD) subject to a large ARL \cite{chan2017optimal,hu2023likelihood}, in which case the aforementioned phase transition when applied to a well-calibrated HC procedure takes the following form. An ``undetectable'' region $t < \Delta^*(r,\beta,\sigma)$ of linear increase of the EDD with the ARL, and a ``detectable'' region $t > \Delta^*(r,\beta,\sigma)$ of a very moderate 
increase of the EDD with a significant increase in ARL. Numerical evaluations in Figure~\ref{fig:just_gap_emp_theo_delay_r0.01beta0.7} below illustrate the two regimes and suggest that the increase in the EDD is logarithmic in the ARL in the detectable region. The logarithmic increase is typical in sequential change-point analysis \cite{poor1998quickest}.

The restriction to monotonic statistics in Theorem~\ref{thm:main_impossible} is sufficient from a practitioner's perspective when seeking a detection procedure that is monotonic in the magnitude of a change gathered across individual streams. However, we would like to remark that a simple extension of similar results in \cite{tony2011optimal} and \cite{kipnis2021logchisquared} appear to yield a lower bound not restricted to such statistics, and we leave such extensions to future work. 

\subsubsection{Relation to previous results}
Typical results in change point detection literature focus on expectations of the stopping time, such as the ARL, $\ex{T \mid H_0}$, and the worst-case EDD, defined by \cite{pollak1985optimal} as $\max_{\tau \geq 1} \mathbb{E}_\tau[T-\tau \mid T \geq \tau]$ when a change occurs at time $\tau$. We address the worst-case EDD in Corollary \ref{cor:HC}. 

The work of  (Chan, 2017) \cite{chan2017optimal} provided the asymptotic EDD under a setting similar to \eqref{eq:normal_problem} with $\sigma=1$ and subject to an ARL lower bound $\gamma$ that increases in $N$. The EDD therein is identical to the one provided by $T_{\HC}$ in Theorem~\ref{thm:HC}, once we adjust the calibration of the change magnitude and the scaling in the ARL lower bound $\gamma$ of \cite{chan2017optimal} to our setup. Consequently, in this case, $T_{\HC}$ attains the optimal detection delay under the change detection model \eqref{eq:normal_problem}. 

Our analysis limits the growth rate of the horizon $\Tcal$ in $N$ and, therefore, our false alarm run length guarantee and its expected value, i.e., the ARL. This is a restriction compared to \cite{chan2017optimal}
that characterized the tradeoff among $\beta$, $r$, and the ARL lower bound $\gamma$ under the scaling $\log(\gamma) \sim N^{\zeta}$ for $\zeta>0$. Note that our setup corresponds to the case $\zeta \to 0$ in \cite{chan2017optimal}; the minimal asymptotic detection delay $\Delta^*(r,\beta,\sigma)$ coincides with that provided in \cite{chan2017optimal} for $\zeta \to 0$.

Under special cases, the work of (Liu et al. 2021) \cite{liu2021minimax} also delivers asymptotic interplays among the number of streams $N$, the control duration $\Tcal$, the magnitude of a change, and the sparsity level. However, their asymptotic scaling is different than 
ours 
and therefore incomparable. Specifically, in the case of severe sparsity $\beta \in (1/2,1)$, the setting of \cite{liu2021minimax} does not lead to a phase transition. On the other hand, it provides interesting interplays for a relatively dense change $\beta \in (0,1/2)$, whereas detection is trivial in this regime in our setting.  

\subsubsection{Heteroscedastic change}
An increase in the post-change variance $\sigma^2$ reduces $\Delta^*(r,\beta,\sigma)$ and thus leads to a quicker detection based on LR and GLR statistics. This situation is well-understood in the context of offline signal detection in \cite{tony2011optimal} and \cite{stoepker2023sparse}.  
When $\sigma^2$ is dramatically smaller or larger than $1$, statistics based on the cumulative sum such as $Y_{n,t}^{\LR}$ and $Y_{n,t}^{\GLR}$ may be much less effective compared to alternatives statistics. For this reason, it is useful to consider $\sigma$ as a nuisance parameter in our setting. 
In particular, the analysis and its conclusions may change dramatically when $\sigma$ is permitted to increase or vanish with $N$. 

\section{Numerical experiments}
\label{sec:experimnets}

This section presents numerical experiments to validate theoretical results and compare the proposed procedure with existing methods. Throughout, we consider data sampled from the model \eqref{eq:data_model}
with $\mu = \sqrt{2r\log N}$, $\sigma^2 = 1$, and
$p= \Prp{n \in I} =N^{-\beta}$. Namely, normal pre- and post-change distributions \eqref{eq:data_model} with $\sigma = 1$, $\mu$ and $p$ are parametrized using $N$, $r$, and $\beta$ as in \eqref{eq:calibration}. 
The HC hyperparameter is fixed at $\alpha_0=0.2$ based on the recommendations in 
(Donoho and Jin, 2015) \cite{donoho2015higher}.

\subsection{Trajectory illustration}

We start by illustrating the detection procedure when there is no change versus when there is a change. Figure~\ref{fig:exact-CUSUM-HC-r0.05-beta0.5} shows the trajectory $\{\HC_t^*\}$ of \eqref{eq:HC_t_star} over $t$ under $H_0$ (no change) and another one under $H_1$ (when the change happens at 2000). Clearly, in this setting, by setting a threshold $b_t = 5$, one can successfully detect the change. Figure~\ref{fig:exact-CUSUM-HC-r0.05-beta0.5-hist}  shows the empirical distribution of simulated values of ${\HC_t^*}$ before and after the change. It is worth mentioning that the empirical distribution appears to be stable after a ``burn-in'' period of about $t= 200$ instances. We only consider in our experiments the detection statistics after the burn-in period.

\subsection{Calibration of threshold} 
\label{sec:calibration}

To estimate the threshold \( b_t = b \) for controlling ARL under \( H_0 \), we employ an indirect Monte Carlo simulation technique used in (Xie and Siegmund 2013) \cite{xie2013sequential}, to reduce computational cost when calibrating the threshold for large target ARL values. This approach is based on the observation that the stopping time under the null hypothesis is approximately exponentially distributed, i.e., \(\Prp{T_\HC \geq t \mid H_0} \approx e^{-\lambda t}\), which can be argued by large-deviation theoretically. Thus, we first estimate \(\Prp{T_\HC \geq t \mid H_0}\) to obtain $\hat\lambda$ by simulating fixed-length sequences, then use the relationship between the probability of hitting the threshold and its expected value to estimate ARL as \(\ex{T_\HC \mid H_0} \approx 1/\hat{\lambda}\). 

\begin{figure}[h!]
\centering 
\includegraphics[width=.45\linewidth]{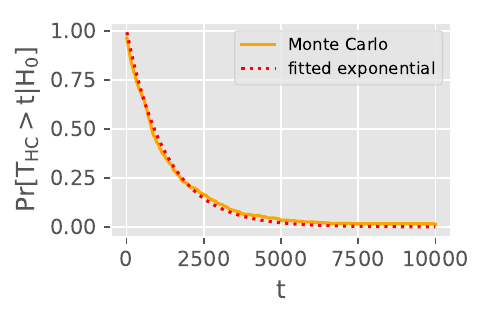}
\caption{
Empirical and fitted survival functions of $T_\HC$ under the null, illustrated for the threshold $b_t = 5$. Data is generated under the null hypothesis with $N=20,000$ across 500 repetitions.}
\label{fig:exp_survival_THC}
\end{figure}

To validate the exponential distribution approximation for the stopping time, we fit the exponential function using least squares; the coefficient of determination exceeds $0.99$. Figure~\ref{fig:exp_survival_THC} shows that the empirical tail aligns closely with the fitted exponential function. We observed similar fits in similar experiments over a range of values besides $b = 5$.

\begin{figure}[t!]
\centering 

\begin{subfigure}[h]{0.7\linewidth}
\includegraphics[width=\linewidth]{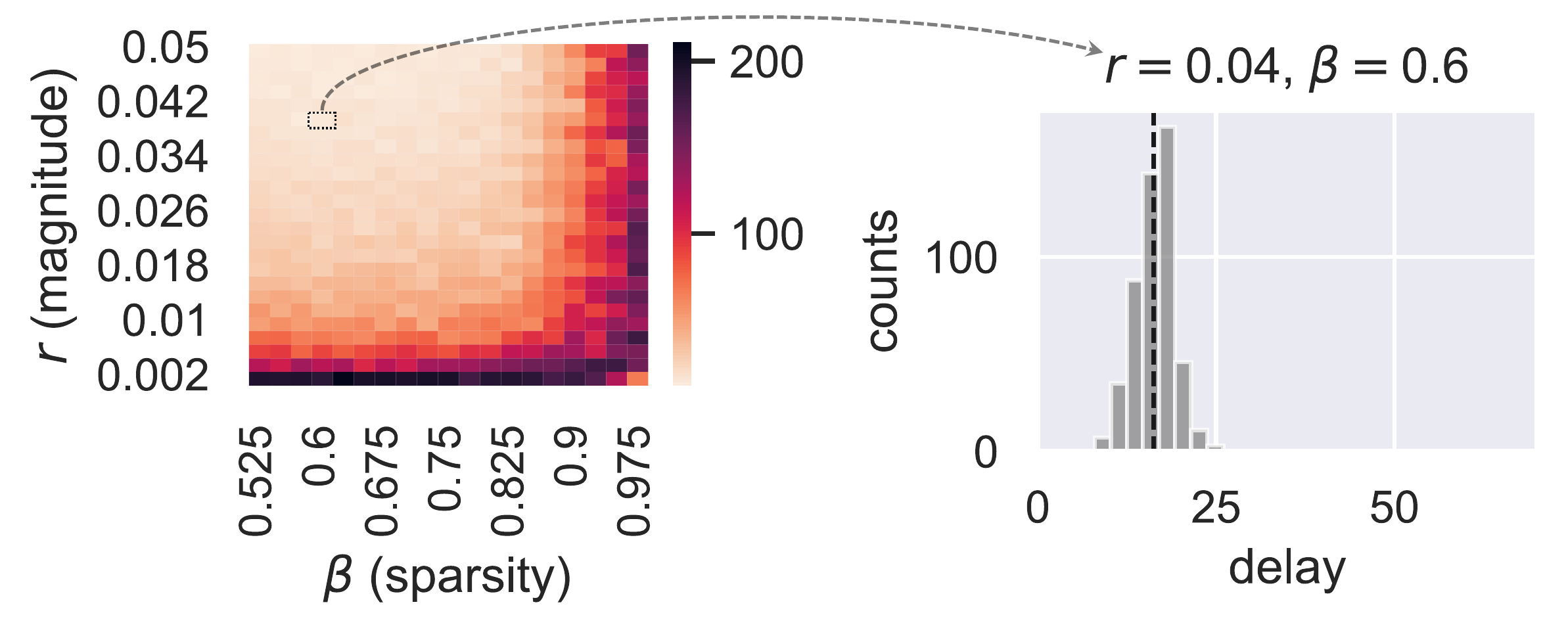}

\label{fig:phase-transition-EDD-sub-rhodivr}
\end{subfigure}
\caption{
Detection delays in sparse normal mean-shift using the HC procedure. Approximately $N^{1-\beta}$ streams undergo a change from $\mathcal{N}(0,1)$ to $\mathcal{N}(\sqrt{2r\log(N)},1)$, with $N = 5,000$ and a grid of $(r,\beta)$ for change magnitude and sparsity. The target ARL is set to be $5,000$. P-values for each stream are computed using the CUSUM statistic. Left: Average detection delay across 500 Monte Carlo trials for each $(r,\beta)$ configuration. Right: Histogram of detection delay for a single $(r,\beta)$ configuration, with the dashed vertical line indicating the corresponding EDD.}
\label{fig:figure4}
\end{figure}

\subsection{Validation of theoretical results}

\subsubsection{EDD dependence on sparsity and change magnitude}

We begin by illustrating the performance of the HC procedure as the sparsity and magnitude of the change vary. Figure~\ref{fig:figure4} shows EDD over a grid of sparsity levels and change magnitudes, with ARL calibrated to be 5000, along with the histogram of detection delays for a single configuration. The EDD and ARL are estimated using 500 Monte Carlo repetitions. The results confirm that the detection delay increases as the change's magnitude becomes weaker and sparser.

\begin{figure}[t!]
\centering 
\begin{subfigure}[h]{0.33\linewidth}
\includegraphics[width=\linewidth]{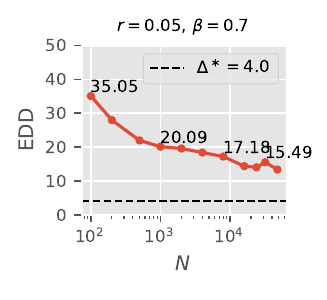}
\caption{}
\end{subfigure}
\begin{subfigure}[h]{0.33\linewidth}
\includegraphics[width=\linewidth]{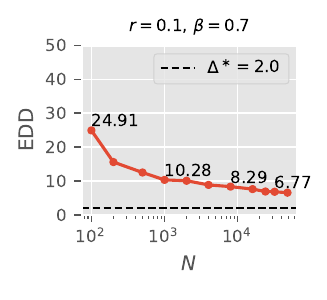}
\caption{}
\end{subfigure}
\begin{subfigure}[h]{0.32\linewidth}
\includegraphics[width=\linewidth]{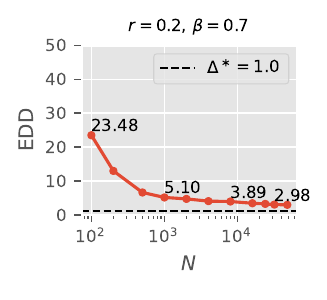}
\caption{}
\end{subfigure}

\begin{subfigure}[h]{0.3\linewidth}
\includegraphics[width=\linewidth]{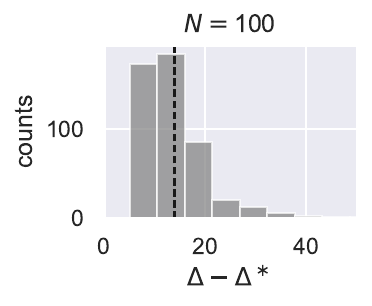}
\caption{}
\end{subfigure}
\begin{subfigure}[h]{0.3\linewidth}
\includegraphics[width=\linewidth]{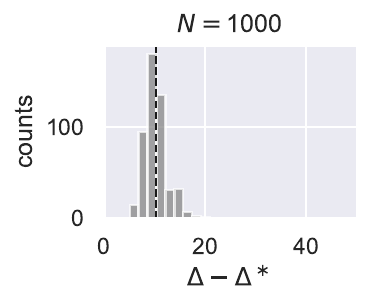}
\caption{}
\end{subfigure}
\begin{subfigure}[h]{0.3\linewidth}
\includegraphics[width=\linewidth]{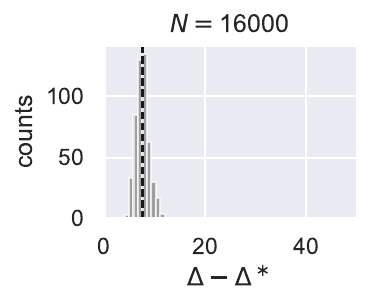}
\caption{}
\end{subfigure}

\caption{Convergence of detection delay $\Delta = T_{\HC}-\tau$ to its theoretical asymptotic value as the number of streams $N$ increases. Data is generated according to \eqref{eq:data_model}, with P-values for each stream computed using the CUSUM statistic. The thresholds are tuned to meet a target ARL $5000$, and results are based on 500 Monte Carlo repetitions. (a)--(c): Expected detection delay $\ex{\Delta \mid \Delta \geq T_{\HC} \geq \tau}$ versus $N$ for $\beta = 0.7$ and $r\in\{0.05,0.1,0.2\}$. Dashed horizontal lines indicate the theoretical asymptotic delay $\Delta^*(r,\beta,\sigma)$. (d)--(f):  Histograms of 
$\Delta-\Delta^\ast(r,\beta, 1)$
for fixed $(r,\beta) = (0.1, 0.7)$ and $N\in\{100,1000,16000\}$; dashed vertical lines represent EDD.}
\label{fig:gap-between-EDD-rho}
\end{figure}

\subsubsection{Convergence of detection delay}

For a finite $N$, there is a gap between the empirical detection delay and the asymptotic detection delay $\Delta^\ast(r,\beta, \sigma)$. As predicted by our asymptotic theory, this gap may shrink with larger $N$ and proper calibration of the threshold $b_t$. The example in this section verified this numerically. 

\begin{figure}[t!]
\centering 
\begin{subfigure}[h]{0.45\linewidth}
\includegraphics[width=\linewidth]{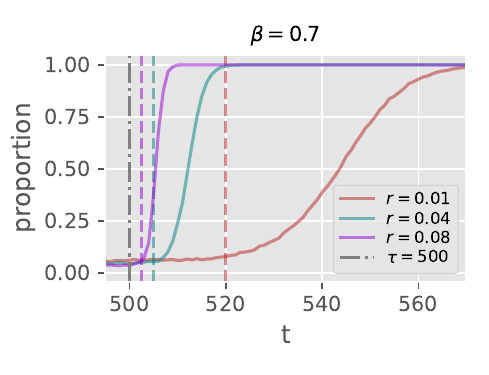}
\label{fig:just_gap_emp_theo_delay_at}
\end{subfigure}

\caption{Rolling empirical probability of detection for HC-procedure over time $t$, computed over 4000 Monte Carlo trials, for various change magnitudes controlled by $r$ via \eqref{eq:mu_r} and $N = 20,000$. The P-value is calculated using the CUSUM statistic at each stream. At each time $t$, the percentage of trajectories with detection statistics exceeding the 95 percentile of $\HC_t^\ast$ under the null hypothesis is shown. Dashed lines indicate $\tau+ \Delta^\ast(r,\beta,1)$, corresponding to the theoretical asymptotic time, exceeding which detection is possible. 
}
\label{fig:just_gap_emp_theo_delay_rListbeta0.7}
\end{figure}

The first row of Figure~\ref{fig:gap-between-EDD-rho} illustrates the convergence of the EDD to $\Delta^\ast(r,\beta, \sigma)$: the gap decreases with $N$ for various change magnitude $r$.  The second row of Figure~\ref{fig:gap-between-EDD-rho} shows that the distribution of the detection delay concentrates around $\Delta^\ast(r,\beta,\sigma)$ as $N$ grows.
Figure~\ref{fig:just_gap_emp_theo_delay_rListbeta0.7} illustrates the probability of detection, over time, subject to an ARL control. This probability increases to one quickly when $t$ exceeds the asymptotic theoretical delay $\Delta^\ast(r,\beta,\sigma)$. 

\subsubsection{Detection delay and run length tradeoff}

\begin{figure}[t!]
\centering 

\begin{subfigure}[h]{0.49\linewidth}
\includegraphics[width=\linewidth]{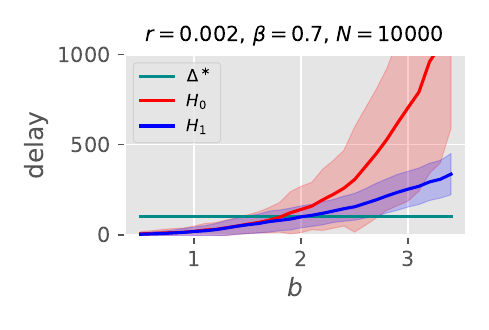}
\caption{}
\label{fig:DD-RL-phasetransition-all}
\end{subfigure}
\begin{subfigure}[h]{0.49\linewidth}
\includegraphics[width=\linewidth]{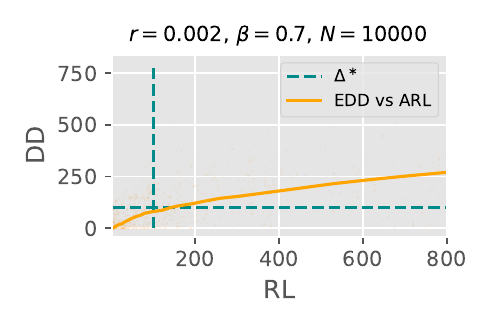}
\caption{}
\label{fig:DD-RL-phasetransition-linear}
\end{subfigure}
\caption{Detection delay (DD) and average no-false-alarm run length (RL) of sparse change point detection using the HC procedure. (a) Empirical expected DD (i.e., EDD) and RL (i.e., ARL) versus the HC threshold $b$. Colored regions represent standard errors. (b) The curve $(\mathrm{ARL}(b), \mathrm{EDD}(b))$ is based on empirical measurements of many HC threshold $b$ values. As $N\to \infty$, Theorems~\ref{thm:main_impossible} and \ref{thm:HC} predicts that the point of significant distinction between the curves in (a) converges to $\Delta^*(r,\beta,\sigma)$, and the bending point in (b) at which the relation between EDD and ARL is no longer linear is at $\Delta^*(r,\beta,1), \Delta^*(r,\beta,1)$. 
The data involves $100$ independent experiments with a change at time $t = 0$. Each experiment uses normal data following \eqref{eq:data_model} with $N = 10,000$ streams, roughly $N^{1-\beta} \approx 16$ affected streams $(\beta=0.7)$ by a mean shift of $\sqrt{2 r \log(N)} \approx 0.2$ ($r=0.002$), and the CUSUM statistic \eqref{eq:YLR_def} in each stream.
}
\label{fig:just_gap_emp_theo_delay_r0.01beta0.7}
\end{figure}

This example illustrates the phase-transition phenomenon discussed in Section~\ref{sec:phase}: an ``undetectable'' regime for $t < \tau+ \Delta^*(r,\beta,\sigma)$, where the expected detection delay (EDD) increases linearly with the ARL, and a ``detectable'' regime for $t \geq \tau+ \Delta^*(r,\beta,\sigma)$, where the EDD increases only moderately despite a significant increase in ARL.
Figure~\ref{fig:just_gap_emp_theo_delay_r0.01beta0.7} illustrates the mean and standard deviation of detection delay (DD) and the RL across multiple trials for varying HC detection thresholds. 
As predicted in Theorem~\ref{thm:main_impossible}, DD and RL are indistinguishable for delays smaller than $\Delta^\ast(r,\beta,\sigma)$. As predicted in Theorem~\ref{thm:HC}, for larger delays, it is feasible to calibrate a procedure based on HC that yields a significantly larger RL relative to DD. Namely, a meaningful detection is possible. 

\subsection{Comparisons with other detection procedures}
\label{subsec:comparisons}

We compare the HC procedure with various existing change point detection procedures, some of which are designed for detecting sparse changes. We examine the procedures' EDD while calibrating each procedure's threshold to reach a target ARL of 5000. For simulating ARL, we set $\mathcal{T} = 20,000$, and the corresponding thresholds are listed in Table~\ref{tab:thresholds_tuning}. The ARLs for each procedure were estimated via Monte Carlo simulations based on 500 trials.
\begin{enumerate}[(i)]
    \item Xie and Siegmund's (XS) \cite{xie2013sequential}: 
\[
T_{\mathrm{XS}}=\inf \left\{t: \max_{t-w\leq k<t} \sum_{n=1}^N \log(1-p_0+p_0 e^{(W_{t,k,n}^+)^2/2}) > b\right\},
\]
where $W_{t,k,n}$ is computed for each stream $n$ via \eqref{eq:YGLR_def}, and $z^+ = \max\{z, 0\}$; here $p_0 = 1/\sqrt{N}$, and $w = 200$.
\item Chan's procedure \cite{chan2017optimal}: 
\[
T_\Chan =\inf \left\{t: \max_{t-w\leq k<t} \sum_{n=1}^N g\left(W_{t,k,n}^+\right) \geq b\right\},
\]
where $g(z)=\log \left[1+p_0\left(C e^{z^2 / 4}-1\right)\right]$ and $C=2(\sqrt{2}-1)$, and we use $p_0 = 1/\sqrt N$, $w = 200$. 
\item Chen, Wang, and Samworth (CWS) \cite{chen2022high}: The work proposes a likelihood ratio-based detection procedure for Gaussian mean-shift detection by testing against simple alternatives of varying signal magnitude scales in each coordinate and aggregates the test statistics across scales and coordinates. Here, for comparison,  we employ the algorithm version from \cite[Sec.~3.3]{chen2022high}, which adapts to both dense and sparse changes. Following the same method as in \cite[Sec.~4.1]{chen2022high}, we tune the three detection thresholds to meet the ARL constraint. The key hyperparameter in this procedure is the lower bound of the $\ell_2$ norm of the vector mean change across all streams; we set this to be 1, matching the lower bound of norms for all possible changes considered in our experiments. 

\item Chen and Chan's (Chen+Chan) \cite{hu2023likelihood}: The work proposes a detection statistic for offline change-point detection, based on a score statistic of P-value distribution departure from the uniform distribution. Here, we modify the statistic for online change-point detection, as follows
\[
T_\CC\left(\lambda_1,\lambda_2\right)=\inf \left\{t: \sum_{n=1}^N \log \left(1+\frac{\lambda_1 \log N}{N} g_1(\pi_{n,t})+\frac{\lambda_2}{\sqrt{N \log N}} g_2(\pi_{n,t})\right) \geq b\right\},
\]
where for $\lambda_1 \geq 0,\, \lambda_2>0$, 
$g_1(z)=\frac{1}{z(2-\log z)^2}-\frac{1}{2}$, and $g_2(z)=\frac{1}{\sqrt{z}}-2$; in the experiment we choose $\lambda_1 = 1$ and $\lambda_2 = \sqrt{\log \Tcal/\log\log \Tcal}$, following the same choices as in \cite{hu2023likelihood}. 
\item logp sum: the procedure combines P-values according to Fisher's method as
\[
T_\Sum =\inf \left\{t: -\sum_{n=1}^N \log(\pi_{n,t}) \geq b\right\}. 
\]
The rationale is to aggregate the P-values across all data streams uniformly; this can work well when all streams are affected by the change, but can be less efficient when the change only affects a few coordinates, as seen in our experiments. 
\item logp min: the procedure uses the smallest P-value as in Bonferroni-type analysis:
\[
T_\Min = \inf \left\{t: -\min_{n=1,\ldots,N} \log(\pi_{n,t}) \geq b\right\}. 
\]
The procedure works well when only one data stream is affected by the change; however, it may lack combining statistical power when more than one stream is affected by the change.  
\item   orderp SSBH: A test inspired by a statistic associated with the 
Seeger-Simes family-wise error controlling
and the Benjamini-Hochberg false discovery rate controlling, \cite{benjamini2001control}, which combines the P-values with weights: 
\[
T_\SSBH = \inf \left\{t: -\min_{n=1,\ldots,N} \frac{\pi_{(n),t}}{n/N} \geq b\right\}. 
\]
\end{enumerate}

\begin{table}[h!]
     \caption{Threshold $b$ for stopping rules corresponding to $\ARL = 5000$. 
     }
    \label{tab:thresholds_tuning}
    \begin{subtable}[h]{1\textwidth}
        \centering
        \begin{scriptsize}
        \begin{tabular}{c cc cc}
        \toprule
        & \multicolumn{2}{c}{$N=100$} & \multicolumn{2}{c}{$N=10^4$}\\
        \cmidrule(lr){2-3} \cmidrule(lr){4-5}\\[-7pt]
        & $b$ & ARL & $b$ & ARL\\
        \cmidrule(lr){2-2} \cmidrule(lr){3-3} \cmidrule(lr){4-4} \cmidrule(lr){5-5}\\[-6pt]
        XS \cite{xie2013sequential}
        & 19.50
        & 4968
        & 113.59
        & 5000   \\[3pt]
        Chan \cite{chan2017optimal}
        & 4.25
        & 5066
        & {4.52} 
        & 5005 \\[3pt] 
        Chen+Chan \cite{hu2023likelihood}
        & 2.89
        & 4992
        & -0.66
        & 5012 \\[3pt]
        logp sum
        & 119.60
        & 4997
        & 9696.15
        & 4999\\[3pt]
        logp min
        & 10.39
        & 4968
        & 14.21
        & 5021 \\[3pt]
        orderp SSBH
        & $-3.06\times 10^{-3}$
        & 5042
        & $-6.77\times 10^{-3}$
        & 5017 \\[3pt]
        HC-WLCUSUM
        & 9.93
        & 4966
        & 12.11
        & 4990  \\
        \bottomrule
		\end{tabular}   
        \end{scriptsize}
    \end{subtable}
\end{table}

\begin{table}[h!]
     \caption{EDD of several detection procedures and values of $|I|$ and $N=100$; data is normal with  $r=1$. EDDs are calculated from 500 repetitions, and the standard errors are in parentheses. The smallest EDD (up to standard errors) is bolded. }
    \label{tab:small_scale_fix_r}
    \begin{subtable}[h]{1\textwidth}
        \centering
        \begin{scriptsize}
        \begin{tabular}{l rr rr rr r}
        \toprule
        & \multicolumn{7}{c}{$|I|$}\\
        \cmidrule(lr){2-8} \\ [-7pt]
        & 1 & 3 & 5 & 10 & 30 & 50 & 100\\
        \midrule
        XS \cite{xie2013sequential}
        & 30.1 (0.55)
        & 13.2 (0.20)
        & 9.3 (0.14)
        & {\bf5.7} (0.08)
        & {2.5} (0.03)
        & {1.8} (0.02)
        & {\bf1.0} (0.01)\\[3pt]
        XS presented in \cite{chan2017optimal}
        & 31.6 & 14.2 & 10.4 & 6.7 & 3.5 & 2.8 & 2.0 \\[3pt]
        Chan \cite{chan2017optimal}
        & 25.7 (0.49)
        & 13.3 (0.2)
        & 9.7 (0.14)
        & 6.3 (0.08)
        & 2.9 (0.04)
        & 2.0 (0.02)
        & 1.1 (0.01)\\[3pt]
        Chan presented in \cite{chan2017optimal} 
        & 26.8 & 13.4 & 9.6 & 6.4 & 2.8 & 2.0 & 1.1 \\[3pt]
        CWS \cite{chen2022high} 
        & 37.2 (0.42)
        & 15.8 (0.21)
        & 10.6 (0.14)
        & 6.5 (0.09)
        & {\bf 2.2} (0.04)
        & {\bf 1.2} (0.02)
        & {\bf 1.0} (0.00) \\[3pt]
        Chen+Chan \cite{hu2023likelihood} 
        & 28.6 (0.50)
        & 15.8 (0.21)
        & 11.5 (0.14)
        & 7.3 (0.08)
        & 3.6 (0.03)
        & 2.5 (0.03)
        & 1.8 (0.02)\\[3pt]
        logp sum
        & 613.0 (11.19)
        & 33.1 (0.48)
        & 20.0 (0.20)
        & 10.6 (0.10)
        & 4.1 (0.03)
        & 2.7 (0.02)
        & 1.8 (0.02) \\[3pt]
        logp min
        & 18.2 (0.38)
        & 12.2 (0.22)
        & 9.8 (0.17)
        & 8.0 (0.13)
        & 5.6 (0.08)
        & 4.7 (0.07)
        & 4.0 (0.05)\\[3pt]
        orderp SSBH
        & 18.2 (0.38)
        & 12.2 (0.22)
        & 9.7 (0.17)
        & 7.9 (0.12)
        & 5.5 (0.08)
        & 4.7 (0.07)
        & 3.9 (0.05)\\[3pt]
        HC-WLCUSUM
        & {\bf16.3} (0.35)
        & {\bf10.3} (0.19)
        & {\bf8.2} (0.14)
        & 6.4 (0.10)
        & 4.0 (0.05)
        & 3.2 (0.04)
        & 2.3 (0.02)\\
        \bottomrule
		\end{tabular}   
        \end{scriptsize}
    \end{subtable}
\end{table}

\begin{table}[h!]
     \caption{
     EDD of several detection procedures and values of $r$, and $N=100$; the data is normal as in \eqref{eq:data_model} with a fixed
     $|I|=5$. EDDs are calculated from 500 repetitions and the standard errors are in parentheses. The smallest EDD (up to standard errors) is bolded.
     }
\label{tab:small_scale_fixed_sparsity}
    \begin{subtable}[h]{1\textwidth}
        \centering
        \resizebox{1\columnwidth}{!}{%
        \begin{tabular}{l rr rr rr rr}
        \toprule
        & \multicolumn{8}{c}{$r$}\\
        \cmidrule(lr){2-9} \\ [-7pt]
        & 0.4 & 0.6 & 0.8 & 1.0 & 1.2 & 1.4 & 1.6 & 1.8\\
        \midrule
        XS \cite{xie2013sequential}
        & 50.2 (0.81)
        & 23.7 (0.33)
        & 13.8 (0.20)
        & 9.3 (0.12)
        & 6.4 (0.09)
        & 5.0 (0.07)
        & 4.0 (0.05)
        & 3.3 (0.04)\\[3pt]
        Chan \cite{chan2017optimal}
        & 53.5 (0.86)
        & 25.0 (0.36)
        & 14.4 (0.21)
        & 9.9 (0.13)
        & 6.8 (0.09)
        & 5.3 (0.07)
        & 4.2 (0.06)
        & 3.5 (0.05)\\[3pt]
        CWS \cite{chen2022high} 
        & 62.9 (0.91)
        & 28.9 (0.40)
        & 16.4 (0.22)
        & 10.7 (0.15)
        & 7.1 (0.10)
        & 5.4 (0.07)
        & 4.1 (0.06)
        & 3.1 (0.05)\\[3pt]
        Chen+Chan \cite{hu2023likelihood} 
        & 63.5 (0.85)
        & 29.5 (0.37)
        & 17.2 (0.23)
        & 11.7 (0.13)
        & 8.1 (0.10)
        & 6.2 (0.07)
        & 5.0 (0.06)
        & 4.2 (0.05)\\[3pt]
         logp sum
        & 105.3 (1.23)
        & 49.5 (0.56)
        & 29.3 (0.34)
        & 20.1 (0.20)
        & 14.4 (0.14)
        & 11.1 (0.1)
        & 8.8 (0.08)
        & 7.4 (0.07) \\[3pt]
        logp min
        & 56.4 (1.08)
        & 25.5 (0.44)
        & 15.2 (0.26)
        & 10.2 (0.17)
        & 7.1 (0.12)
        & 5.5 (0.09)
        & 4.4 (0.07)
        & 3.6 (0.05) \\[3pt]
        orderp SSBH
        & 56.2 (1.07)
        & 25.3 (0.44)
        & 15.1 (0.26)
        & 10.1 (0.16)
        & 7.0 (0.12)
        & 5.5 (0.08)
        & 4.4 (0.07)
        & 3.6 (0.05)\\[3pt]
        HC-WLCUSUM
        & {\bf44.9} (0.89)
        & {\bf20.9} (0.37)
        & {\bf12.6} (0.22)
        & {\bf8.3} (0.13)
        & {\bf5.8} (0.10)
        & {\bf4.6} (0.07)
        & {\bf3.7} (0.06)
        & {\bf3.0} (0.04)\\
        \bottomrule
		\end{tabular}   }
    \end{subtable}
\end{table}

\begin{table}[t!]
     \caption{Large $N$ comparsion: 
     EDD for varying $|I|$, while fixing $r=1$ and $N=10^4$. EDDs are calculated from 500 repetitions, and the standard errors are in parentheses. In each setting, the smallest EDD (up to standard errors) is bolded. ``--'' denotes ``failure to detect'': the procedure did not raise an alarm by the end of the detection time horizon 1000 for all sequences. ``N/A" means that a specific result is not available.}
    \label{tab:large_scale_fix_r}
    \begin{subtable}[h]{1\textwidth}
        \centering
        \begin{scriptsize}
        \begin{tabular}{c cc cc cc}
        \toprule
        & \multicolumn{6}{c}{$|I|$}\\
        \cmidrule(lr){2-7} \\ [-7pt]
        & $1$ & $5$ & $10$ & $10^2$ & $10^3$ & $10^4$\\
        \midrule
        XS \cite{xie2013sequential}
        & 64.5 (0.9)
        & 20.8 (0.22)
        & 13.9 (0.12)
        & {\bf 4.0} (0.03)
        & {\bf 1.0} (0.0)
        & {\bf 1.0} (0.0)\\[3pt]
        Chan \cite{chan2017optimal}
        & 37.3 (0.54)
        & 18.0 (0.2)
        & 13.3 (0.13)
        & 4.5 (0.04)
        & {\bf 1.0} (0.0)
        & {\bf 1.0} (0.0)\\[3pt]
        Chan presented in \cite{chan2017optimal} 
        & 37.7 
        & N/A
        & 13.3 
        & 4.5 
        & {\bf1.0} 
        & {\bf1.0}\\[3pt]
        CWS \cite{chen2022high} 
        & 64.4 (0.61)
        & 20.0 (0.20)
        & 14.4 (0.13)
        & 5.9 (0.05)
        & {\bf 1.0} (0.0)
        & {\bf 1.0} (0.0)\\[3pt]
        Chen+Chan \cite{hu2023likelihood} 
        & 870.9 (5.67)
        & 35.2 (0.31)
        & 23.5 (0.16)
        & 9.6 (0.05)
        & 3.7 (0.02)
        & {\bf 1.0} (0.0)\\[3pt]
        logp sum
        & --
        & 999.9 (0.07)
        & 998.1 (0.59)
        & 263.4 (0.79)
        & 7.2 (0.02)
        & {\bf 1.0} (0.0) \\[3pt]
        logp min
        & {\bf 25.8} (0.47)
        & {\bf 15.5} (0.24)
        & 12.7 (0.18)
        & 7.2 (0.09)
        & 4.2 (0.05)
        & 2.8 (0.03)\\[3pt]
        orderp SSBH
        & {\bf 25.8} (0.47)
        & {\bf 15.5} (0.24)
        & 12.6 (0.18)
        & 7.1 (0.09)
        & 4.2 (0.05)
        & 2.7 (0.03) \\[3pt]
        HC-WLCUSUM
        & {\bf 25.8} (0.47)
        & {\bf 15.3} (0.23)
        & {\bf 12.4} (0.17)
        & 6.7 (0.08)
        & 3.5 (0.03)
        & {\bf 1.0} (0.0)\\
        \bottomrule
		\end{tabular}   
        \end{scriptsize}
    \end{subtable}
\end{table}

\begin{table}[t!]
     \caption{Large $N$ comparison: EDD for varying $r$, while fixing $|I| = 10$, and $N=10^4$. EDDs are calculated from 500 repetitions, and the standard errors are in parentheses. The smallest EDD (up to standard errors) is bolded in each setting. 
     }
    \label{tab:large_scale_fix_I}
    \begin{subtable}[h]{1\textwidth}
        \centering
        \resizebox{1\columnwidth}{!}{%
        \begin{tabular}{c cc cc cc cc}
        \toprule
        & \multicolumn{8}{c}{$r$}\\
        \cmidrule(lr){2-9} \\ [-7pt]
        & $0.4$ & $0.6$ & $0.8$ & $1.0$ & $1.2$ & $1.4$ & $1.6$ & $1.8$\\
        \midrule
        XS \cite{xie2013sequential}
        & 79.7 (0.88)
        & 36.7 (0.4)
        & 21.2 (0.21)
        & 13.8 (0.13)
        & 9.9 (0.09)
        & 7.4 (0.07)
        & 5.8 (0.05)
        & 4.7 (0.04)\\[3pt]
        Chan \cite{chan2017optimal}
        & 76.9 (0.92)
        & 35.3 (0.41)
        & 20.5 (0.21)
        & 13.4 (0.13)
        & 9.4 (0.09)
        & 7.1 (0.07)
        & 5.6 (0.05)
        & 4.6 (0.04)\\[3pt]
        CWS \cite{chen2022high} 
        & 87.2 (0.84)
        & 39.5 (0.35)
        & 22.5 (0.20)
        & 14.3 (0.14)
        & 9.9 (0.09)
        & 7.3 (0.07)
        & 5.5 (0.05)
        & {\bf 4.3} (0.04)\\[3pt]
        Chen+Chan \cite{hu2023likelihood} 
        & 135.2 (1.05)
        & 61.5 (0.49)
        & 36.0 (0.28)
        & 23.5 (0.17)
        & 16.8 (0.11)
        & 12.7 (0.09)
        & 9.9 (0.06)
        & 8.1 (0.06)\\[3pt]
        logp sum
        & 997.9 (0.54)
        & 997.2 (0.73)
        & 997.4 (0.83)
        & 997.6 (0.66)
        & 997.9 (0.68)
        & 997.4 (0.62)
        & 998.0 (0.58)
        & 998.3 (0.49) \\[3pt]
        logp min
        & {\bf 70.7} (1.23)
        & 32.4 (0.55)
        & 19.3 (0.29)
        & 12.6 (0.19)
        & 9.0 (0.13)
        & 6.9 (0.09)
        & 5.4 (0.07)
        & 4.4 (0.06)\\[3pt]
        orderp SSBH
        & {\bf 70.5} (1.22)
        & {\bf 32.3} (0.55)
        & 19.3 (0.29)
        & 12.6 (0.19)
        & 9.0 (0.13)
        & 6.8 (0.09)
        & 5.4 (0.07)
        & 4.4 (0.06)\\[3pt]
        HC-WLCUSUM
        & {\bf 69.7} (1.19)
        & {\bf 31.8} (0.53)
        & {\bf 18.9} (0.28)
        & {\bf 12.3} (0.18)
        & {\bf 8.8} (0.12)
        & {\bf 6.7} (0.09)
        & {\bf 5.3} (0.07)
        & {\bf 4.3} (0.05)\\
        \bottomrule
		\end{tabular}   }
    \end{subtable}
\end{table}

To ensure a fair comparison, we tune the thresholds $b$ for each procedure to achieve the same target ARL of $5,000$. The thresholds, determined through simulation over 500 repetitions, are provided in Table~\ref{tab:thresholds_tuning}. We evaluate the methods under the following settings for moderate and high dimensions: (i) $N = 100$, $r = 1$, with varying $|I|$; (ii) $N = 100$, $|I| = 5$, with varying $r$; (iii) $N = 10^4$, $r = 1$, with varying $|I|$; and (iv) $N = 10^4$, $|I| = 10$, with varying $r$.

For $N = 100$, the results are shown in Tables~\ref{tab:small_scale_fix_r} (for varying $|I|$ with fixed $r$) and \ref{tab:small_scale_fixed_sparsity} (for varying $r$ with fixed $|I|$). In setting (i), the HC-based procedure achieves the smallest EDDs for relatively sparse shifts ($|I| = 1, 3, 5$), while the XS and CWS procedures perform better in denser cases. In setting (ii), where the sparsity is fixed at $|I| = 5$ and the shift magnitude $r$ varies, the HC-based procedure consistently outperforms other methods across different values of $r$ for sparse shifts.

For larger-scale experiments with $N = 10^4$, the results are presented in Tables~\ref{tab:large_scale_fix_r} (for varying $|I|$ with fixed $r$)  and \ref{tab:large_scale_fix_I} (for varying $r$ with fixed $|I|$). The HC-based procedure achieves the smallest EDDs for relatively sparse shifts ($|I| = 1, 5, 10$). In Table~\ref{tab:large_scale_fix_I}, with fixed sparsity $|I| = 10$ and varying $r$, the HC-based procedure consistently performs well. As a sanity check, Tables~\ref{tab:small_scale_fix_r} and \ref{tab:large_scale_fix_r} also include results for the XS and Chan methods from \cite{chan2017optimal}, which align closely with our reproduced results.

\vspace{.1in}
\noindent{\it Computational complexity with respect to dimension $N$.} 
Here, the complexity corresponds to the computation of the detection statistics at each time $t$. Due to the requirement to sort the P-values, the HC procedure and procedures (vi) and (vii) have a time complexity of $O(N\log N)$. The CWS procedure has a complexity of $O(N^2 \log N)$ because it aggregates statistics across every pair of streams and considers up to $\log(N)$ different change magnitudes. All other baseline methods have a complexity of $O(N)$.

\section{Proofs}\label{sec:proof}
We now present the proof of the main result. In Section \ref{sec:background}, we provide the necessary background on offline hypothesis testing for rare and weak moderately departure models and asymptotically log-chi-squared P-values. In Section~\ref{sec:multistream_RMD} we apply these previous results to sparse sequential change point detection with asymptotically log-chi-squared P-values. 
Section \ref{sec:P_value} shows that the P-values relevant to this paper satisfy these properties. Section \ref{sec:main_result_proof} proves the main result.

\subsection{Rare moderately departing P-values}\label{sec:background}

Rare (sparse) and weak models for signal detection involve detection or classification challenges in which most of the features are useless (pure noise), except perhaps very few features. The locations of the useful features, if there are any, are unknown to us in advance \cite{donoho2004higher,jin2016rare}. As explained in \cite{kipnis2021logchisquared}, previously studied rare and weak models in which departures of non-null features are on the moderate deviation scale can be carried out under the following testing problem, involving independent P-values $\{\pi_n\}_{n=1}^N$ from individual features. 
\begin{align}
\begin{split}
    H_0^{(N)} \, & :\, \pi_n \sim \Unif(0,1),\quad \forall n=1,\ldots,N, \\
    H_1^{(N)} \, & :\, \pi_n \sim (1-p)\Unif(0,1) + p Q_n^{(N)}, \quad \forall n=1,\ldots,N,
    \label{eq:hyp_test_pvalues}
    \end{split}
\end{align}
where the sequence of distributions $\{Q_n^{(N)}\}_{n=1}^N$ obey:
\begin{align}
X_n \simiid Q_n^{(N)} \Leftrightarrow  \lim_{N \to \infty} \max_{n=1,\ldots,N} \abs{ \frac{- \log \Prp{ -2\log X_n \geq 2 q \log(N)} }{\log(N)} - \left(\frac{\sqrt{q}-\sqrt{\rho}}{\sigma}\right)^2
} = 0,
\label{eq:logchisqaured_appx}
\end{align}
for all $q> \rho> 0$. 
A sufficient condition for the validity of \eqref{eq:logchisqaured_appx} is that 
\[
-2 \log Q_n^{(N)} \overset{D}{=}  \left( \sqrt{2 \rho \log(N)} + \sigma Z \right)^2(1+o_p(1)),\quad Z \sim \Ncal(0,1),
\]
where $o_p(1)$ indicates a sequence tending to zero in probability as $N \to \infty$ uniformly in $n$. Indeed, if $q > \rho$, for $N$ large enough
\begin{align}
&\Prp{ \left(\sqrt{2 \rho \log(N)} + \sigma Z \right)^2(1+o_p(1)) \geq 2 q \log (N)} \nonumber \\& = 2\Prp{  \frac{(\sqrt{2 q } - \sqrt{2 \rho})\sqrt{\log(N)}}{\sigma}(1+o(1)) \leq Z } \nonumber  \\
& = N^{-\left(\frac{\sqrt{q}-\sqrt{\rho}}{\sigma}\right)^2+o(1)}, \nonumber
\end{align}
the last transition by Mill's ratio \cite{shorack2009empirical}. 

The name ``asymptotically log-chisquared" is because \eqref{eq:logchisqaured_appx} holds in particular when 
\[
X_n \overset{D}{=} \exp\left\{-\frac{1}{2}\left( \sqrt{2 \rho \log(N)} + \sigma Z \right)^2 \right\}. 
\]
Namely, when the $X_n$'s follow a non-central log-chisquared distribution over one degree of freedom, as explained in \cite{kipnis2021logchisquared}, models involving moderate departures in individual features falling under the formulation \eqref{eq:logchisqaured_appx} share many common properties concerning the asymptotic behavior of testing procedures. Two properties relevant to our situation are the asymptotic powerlessness of any global test based on $\pi_{1}\leq \cdots \leq \pi_{N}$ when $\rho < \rho^*(\beta, \sigma)$ and the asymptotic powerfulness of $\HC^*$ of \eqref{eq:HC_star} for $\rho > \rho^*(\beta, \sigma)$. 
\begin{theorem}{\cite[Thms. 1 \& 2]{kipnis2021logchisquared}}
\label{thm:Kipnis2021}
    Fix $\rho >0$, $\sigma>0$, and $\beta >0$. Assume that $p = N^{-\beta}$ and that $\{\pi_n\}_{n=1}^N$ follow \eqref{eq:hyp_test_pvalues}. If $\rho > \rho^*(\beta,\sigma)$, there exists a sequence $\{b^{(N)},\,N=1,2,\ldots\}$ such that 
    \begin{align}
        \label{thm:Kipnis2021_possible}
    \lim_{N \to \infty} \left( \Prp{\HC^* < b^{(N)} \mid H_1^{(N)}} + \Prp{\HC^* \geq b^{(N)} \mid H_0^{(N)} } \right) = 0. 
    \end{align}
    If $\rho < \rho^*(\beta,\sigma)$, for any sequences of test statistics $U = U(\pi_1,\ldots,\pi_N)$ and thresholds $\{b^{(N)},\,N=1,2,\ldots\}$, 
    \begin{align}
    \label{thm:Kipnis2021_impossible}
    \liminf_{N \to \infty} \left( \Prp{U < b^{(N)} \mid H_1^{(N)}} + \Prp{U \geq b^{(N)} \mid H_0^{(N)}} \right)= 1.
    \end{align}
\end{theorem}

\subsection{Multistream change-point detection with rare, moderately departed streams}
\label{sec:multistream_RMD}

We now provide two corollaries of Theorem~\ref{thm:Kipnis2021} when applied to detect a global change point sequentially across multiple streams of moderately departed P-values. Later, we consider the case where each stream is obtained by individually testing each data stream for a change. The proofs of both corollaries are based on triangular array arguments and are provided in Appendix~\ref{sec:proofs}. 

The setting below supposes a random set $I$ such that $ \Prp{n \in I}=p = N^{-\beta}  $, $\beta \in (0,1)$,  independently for all $n=1,\ldots,N$, and a stopping rule based on P-values
\begin{subequations}
\label{eq:p_value_rare_model}
\begin{align}
\label{eq:p_value_rare_model_pval}
    \pi_{n,t} \sim \begin{cases}
        \Unif(0,1), & n\notin I \text{ or } t=1,\ldots,\tau-1, \\
        Q_{n,t}^{(N)},& n \in I\,\text{and}\,\,t= \tau,\ldots,\Tcal.
    \end{cases}
\end{align}
Here the array of distributions $\{Q_{n,t}^{(N)},\,n=1,\ldots,N,\, t=1,\ldots \Tcal \}$ has the property that for every fixed $t$, 
if $X_{n,t} \sim Q_{n,t}^{(N)}$ independently in $n=1,\ldots,N$, then for some $\rho(\Delta) \geq 0$ we have
\begin{align}
\lim_{N \to \infty} \max_{n=1,\ldots,N} \abs{ \frac{- \log \Prp{ -2\log X_{n,t} \geq 2 q \log(N)} }{\log(N)} - \left(\frac{\sqrt{q}-\sqrt{\rho(\Delta)}}{\sigma}\right)^2
} = 0,
\label{eq:logchisqaured_appx_time}
\end{align}
\end{subequations}
for all $q> \rho(\Delta)$. Later on, we show that P-values of LR and GLR tests in the normal heteroscedastic change-point setting  \eqref{eq:normal_problem} with $\mu = \mu_N(r) = \sqrt{2r \log(N)}$ satisfy \eqref{eq:logchisqaured_appx_time} with $\rho(\Delta) = \max\{r(\Delta-\tau+1),0\}$. 

The first corollary says that when $\rho(\Delta) < \rho^*(\beta,\sigma)$, it is impossible to provide a non-trivial detection procedure using any statistics based on $\{\pi_{n,t}\}$ of \eqref{eq:p_value_rare_model_pval} that rejects the global null for large values, i.e., small values of $\{\pi_{n,t}\}$. 
\begin{corollary}
\label{cor:impossible_general}
Define the set $\Theta_{0} := \{\Delta\,:\,\rho(\Delta) \leq \rho^-\}$, for some $\rho^- < \rho^*(\beta,\sigma)$. Namely, $\Theta_{0}$ is the set of ``insufficient'' time delays. Let $T_U$  be a stopping rule based on $U_t := U(\pi_{1,t},\ldots,\pi_{N,t})$ and an array of threshold $\{b_t^{(N)},\,t=1,2,\ldots,\, N=1,2,\ldots\}$. Suppose that  $U(x_1,\ldots,x_N) \in \mathbb R$ is non-increasing in either argument. Then
\begin{align*}
\lim_{N \to \infty} 
\max_{\Delta \in \Theta_0}
\left| \Prp{ T_U \leq \Delta \mid H_1^{(N)}} - \Prp{ T_U \leq \Delta \mid H_0^{(N)}} \right| = 0.
\end{align*}
\end{corollary}

The next corollary says that when $\rho(\Delta) > \rho^*(\beta,\sigma)$, there exists a detection procedure based on higher criticism of P-values obeying \eqref{eq:p_value_rare_model} that asymptotically has full power.
\begin{corollary}
    \label{cor:possible_general}
    Define the set $\Theta_1 := \{\Delta\,: \rho(\Delta) \geq \rho^+\}$ for some $\rho^+ > \rho^*(\beta, \sigma)$. Let $\HC_t^*$ be higher criticism of $\{\pi_{1,t}, \ldots, \pi_{N,t}\}$ of \eqref{eq:p_value_rare_model}. 
    Suppose that $\Tcal = o(\log(N))$. There exists an array of thresholds $\{b_t^{(N)},\, t=1,\ldots,\Tcal,\, N = 1,2\ldots \}$ such that the stopping time $T_{\HC} = \inf \{ t\,:\, \HC^*_t > b_t^{(N)}\}$ satisfies
\begin{align*}
\max_{\Delta \in \Theta_1} \Prp{T_{\HC} > \Delta \mid H_1^{(N)}} \to 0,
\end{align*}
while
\begin{align*}
 \Prp{T_{\HC} \leq \Tcal \mid H_0^{(N)}} \to 0.
\end{align*}
\end{corollary}

\subsection{Log-chisquared properties of change-point detection tests}\label{sec:P_value}

We provide conditions under which the P-values of the likelihood and generalized likelihood ratios are asymptotically log-chisquared uniformly in $t$. These conditions coincide with the asymptotic calibration of the intensity of individual change points and their sparsity to $N$ in \cite{chan2017optimal,hu2023likelihood}. 
One notable difference from \cite{chan2017optimal} is the lower bound on the ARL $\gamma$, which we assume to go to infinity but at a rate arbitrarily small. This is a weaker result than the rate of $\log(\gamma) N^{-\zeta} \to 0$ provided in \cite{chan2017optimal}. 

Recall that for the CUSUM and GLR procedures, monitoring for a change in the data's distribution based on $Y_t^{\LR}$ or $Y_t^{\GLR}$ involves the stopping times
\[
T_{x}^{\LR} := \inf\{ t \,:\, Y_t^{\LR} \geq x \},
\]
and 
\[
T_{x}^{\GLR} := \inf\{ t \,:\, Y_t^{\GLR} \geq x \}.
\]
Recall that the P-values $\pi_t^{\LR}(x)$ and $\pi_t^{\GLR}(x)$ the survival functions of the statistics $Y_t^{\LR}$ and $Y_t^{\GLR}$ under the null. If we plug in the corresponding observed statistics value to replace $x$, we get P-values under the model \eqref{eq:hyp_test_pvalues}.

The distribution of $Y_t^{\square}$, $\square \in \{\LR, \GLR\}$ is continuous under either $H_0$ or $H_1$. Therefore, under $H_0$, we have
\begin{align}
    \pi_{t}^\square (Y_t^\square) \sim \Unif(0,1),\qquad t=1,\ldots,\Tcal.
\end{align}

Under the alternative, the following theorem states that P-values obtained from individual LR or GLR tests in streams experiencing a change under \eqref{eq:normal_problem} are asymptotically log-chisquared with scaling parameter $\sigma$ and non-centrality parameter $\rho = \rho(\Delta) = r[t-\tau + 1]_+$, where $[x]_+ = \max\{0,x\}$. 

\begin{theorem}
    \label{thm:alternative}
Assume that $\mu_N(r) = \sqrt{2 r \log(N)}$ for some $r>0$, and for some $\tau\geq 1$ we have $X_1,\ldots,X_{\tau-1} \simiid \Ncal(0,1)$, $X_\tau,\ldots,X_{\Tcal} \simiid \Ncal(\mu_N(r), \sigma^2)$, and  $\Tcal = o(\log(N))$. Set $\rho(\Delta):= r[t-\tau+1]_+$. For $\square \in \{\LR, \GLR\}$ we have
\begin{align}
    \label{eq:Vt_under_H1}
       -2\log \pi_t^{\square}(Y_t^{\square}) =  \left(\sigma Z + \mu_N(\rho(\Delta)) \right)^2 (1 + o_p(1)),
    \end{align}
where $o_p(1)$ represents a random variable that goes to $0$ in probability as $N \to \infty$, uniformly in $t$ and $\tau$.
\end{theorem}

\subsection{Proof of main results}\label{sec:main_result_proof}

Now we will put everything together to prove the main result for the detection delay in multi-stream with sparse, moderately large change points. 

Recall that we consider the change-point detection problem \eqref{eq:normal_problem} with P-values $\pi_{n,t}^\square$ of \eqref{eq:p-val_def_LR} for $\square = \LR$ and \eqref{eq:p-val_def_GLR} for $\square = \GLR$. These P-values obey \eqref{eq:p_value_rare_model_pval} and, by Theorem~\ref{thm:alternative}, every $Q_{n,t}^{(N)}$ satisfies \eqref{eq:logchisqaured_appx_time} with  $\rho(\Delta) = r[\tau - \Delta + 1]_+$, hence Corollaries~\ref{cor:impossible_general} and \ref{cor:possible_general} apply.  

\begin{proof}[Proof of Theorem~\ref{thm:main_impossible}]
Let $\Delta_0 := \lceil \rho^*(\beta,\sigma)/r \rceil - 1 < \rho^*(\beta,\sigma)/r$, hence 
$r \Delta_0 < \rho^*(\beta,\sigma)$. 
We use Corollary~\ref{cor:impossible_general} with $\rho^- = r[\tau - \Delta_0+1]_+ $ and  $\rho(\Delta) = r[\tau - \Delta + 1]_+$. Thus, any $\Delta \leq \Delta_0 < \rho^*(\beta,\sigma)/r$ is in the set $\Theta_0 = \{ \Delta\, :\, \rho(\Delta) \leq \rho^- \}$. Theorem~\ref{thm:main_impossible} follows. 
\end{proof}

\begin{proof}[Proof of Theorem \ref{thm:HC}]
Let $\Delta^*$ be the minimal $\Delta$ such that $\rho(\Delta) = r[\Delta-\tau+1]$ (strictly) exceeds $\rho^*(\beta,\sigma)$. 
Because $\rho^*(\beta,\sigma)/r$ is not an integer, 
$\Delta^* - \tau + 1 = \lceil \rho^*(\beta,\sigma)/r \rceil $, and thus $\rho(\Delta) \geq \rho(\Delta^*) > \rho^*(\beta,\sigma)/r$ for all $t \geq t^*$. Use Corollary~\ref{cor:possible_general} with $\rho^+ = r \Delta^*$, thus
$\Theta_1 = \{\Delta \geq \Delta^* \}$. It follows that there exists an array of thresholds such that 
\begin{align}
    \label{eq:thm:main:proof1}
\Prp{T_{\HC} \leq \Delta^* \mid H_1^{(N)}} \to 1,
\end{align}
while 
\[
\Prp{T_{\HC} \leq \Tcal \mid H_0^{(N)}} \to 0.
\]
On the other hand, because $\HC_t^*$ is non-decreasing as any of $\pi_{1,t},\ldots,\pi_{N,t}$ is decreasing, when $\Delta^*-1< \rho^*(\beta,\sigma)/r$, Corollary~\ref{cor:impossible_general} implies that 
\begin{align}
    \label{eq:thm:main:proof2}
    \Prp{T_{\HC} \leq \Delta^*-1 \mid H_1^{(N)}} \to 0.
\end{align}
Equations \eqref{eq:thm:main:proof1} and \eqref{eq:thm:main:proof2} implies 
\[
\Prp{T_{\HC} = \Delta^* \mid H_1^{(N)}} \to 1.
\]
Theorem~\ref{thm:HC} follows. 
\end{proof}

\section{Conclusions and open challenges}\label{sec:discussion}

This paper introduced a change-point detection procedure that combines information from individual streams using higher criticism (HC). The method is advantageous for practitioners because higher criticism is a well-understood and popular tool for sparse signal detection, having, among other advantages, a built-in selection mechanism to identify a useful set of suspected departing streams \cite{donoho2015higher}. Extensive numerical evaluations show that the HC-based procedure typically outperforms other methods in simulation studies. Finally, the detection delay of the HC procedure has interesting optimality properties: in a sparse heteroscedastic normal change-point setting with CUSUM or general likelihood ratio tests, this delay converges to the theoretical minimal delay, allowing useful change-point monitoring. 

\vspace{.1in}
Below, we discuss several limitations of our work that invite further exploration.

\vspace{.1in}
\noindent{\it Finer asymptotic delay characterization.}
Our main results show that there exists a detection procedure based on HC whose asymptotic detection delay converges in distribution to $\Delta^*(r,\beta,\sigma)$ as the number of streams $N$ goes to infinity. However, Figure~\ref{fig:gap-between-EDD-rho} suggests that this convergence is relatively slow, hence the limiting detection delay may not provide enough information on the performance for finite $N$. Such information can be provided by characterizing the limiting distribution of the delay, or at least the behavior of higher-order asymptotic terms of its mean.


\vspace{.1in}
\noindent{\it Better ARL guarantee.}
Our theoretical analysis allows the ARL to diverge with $N$ but limits it to 
$o(\log(N))$. Therefore, it provides a lower ARL guarantee than studies such as \cite{chan2017optimal} and \cite{hu2023likelihood} which consider ARL scaling as $\sim e^{N^\zeta}$, $\zeta \in (0,1)$. Despite this limitation of our theoretical analysis, numerical experiments show that detection based on HC performs well compared to other methods under arbitrarily large ARL requirements. Therefore, extending our results to ARL scaling better than $o(\log(N))$ appears to be possible and, thus, an interesting avenue for future work. 
%

\vspace{.1in}
\noindent{\it Information theoretic delay.}
We did not fully characterize the information-theoretic detection delay, as alternative tests applied to individual streams might perform better. For instance, the likelihood ratio (LR) test assumes a known post-change mean but not the post-change variance. 

\vspace{.1in}
\noindent{\it Other sparse models.}
Another direction for future work is the consideration of other sparse data models. For example, the work of \cite{kipnis2021logchisquared} suggests that the most interesting tradeoff between change intensity and sparsity occurs when changes follow a moderate deviation scale in $N$. This insight suggests that sparse sequential change-point detection based on HC has optimality properties under various data models, with affected streams experiencing a moderate change in the distribution. It is also interesting to explore sparse sequential change-point detection under non-moderate distribution changes in analogy to the offline change detection considered in \cite{arias2015sparse,jin2016rare,arias2019detection,arias2020dependence,DonohoKipnis2020}.


\begin{funding}
This work of Tingnan Gong and Yao Xie is partially supported by NSF DMS-2134037, DMS-2220495, and the Coca-Cola Foundation. The work of Alon Kipnis is partially funded by the US-Israel Binational Science Foundation (BSF; grant no. 2022124).
\end{funding}

\bibliographystyle{imsart-number} 
\bibliography{main}       

\begin{thebibliography}{42}

\bibitem{arias2020dependence}
\begin{barticle}[author]
\bauthor{\bsnm{Arias-Castro},~\bfnm{Ery}\binits{E.}},
  \bauthor{\bsnm{Huang},~\bfnm{Rong}\binits{R.}} \AND
  \bauthor{\bsnm{Verzelen},~\bfnm{Nicolas}\binits{N.}}
(\byear{2020}).
\btitle{{Detection of sparse positive dependence}}.
\bjournal{Electronic Journal of Statistics}
\bvolume{14}
\bpages{702 -- 730}.
\end{barticle}
\endbibitem

\bibitem{arias2015sparse}
\begin{barticle}[author]
\bauthor{\bsnm{Arias-Castro},~\bfnm{Ery}\binits{E.}} \AND
  \bauthor{\bsnm{Wang},~\bfnm{Meng}\binits{M.}}
(\byear{2015}).
\btitle{The sparse {P}oisson means model}.
\bjournal{Electronic Journal of Statistics}
\bvolume{9}
\bpages{2170--2201}.
\end{barticle}
\endbibitem

\bibitem{arias2019detection}
\begin{barticle}[author]
\bauthor{\bsnm{Arias-Castro},~\bfnm{Ery}\binits{E.}} \AND
  \bauthor{\bsnm{Ying},~\bfnm{Andrew}\binits{A.}}
(\byear{2019}).
\btitle{Detection of sparse mixtures: Higher criticism and scan statistic}.
\bjournal{Electronic Journal of Statistics}
\bvolume{13}
\bpages{208--230}.
\end{barticle}
\endbibitem

\bibitem{benjamini2001control}
\begin{barticle}[author]
\bauthor{\bsnm{Benjamini},~\bfnm{Yoav}\binits{Y.}} \AND
  \bauthor{\bsnm{Yekutieli},~\bfnm{Daniel}\binits{D.}}
(\byear{2001}).
\btitle{{The control of the false discovery rate in multiple testing under
  dependency}}.
\bjournal{The Annals of Statistics}
\bvolume{29}
\bpages{1165 -- 1188}.
\bdoi{10.1214/aos/1013699998}
\end{barticle}
\endbibitem

\bibitem{bhamidi2018change}
\begin{barticle}[author]
\bauthor{\bsnm{Bhamidi},~\bfnm{Shankar}\binits{S.}},
  \bauthor{\bsnm{Jin},~\bfnm{Jimmy}\binits{J.}} \AND
  \bauthor{\bsnm{Nobel},~\bfnm{Andrew}\binits{A.}}
(\byear{2018}).
\btitle{Change point detection in network models: Preferential attachment and
  long range dependence}.
\bjournal{The Annals of Applied Probability}
\bvolume{28}
\bpages{35--78}.
\end{barticle}
\endbibitem

\bibitem{chan2017optimal}
\begin{barticle}[author]
\bauthor{\bsnm{Chan},~\bfnm{Hock~Peng}\binits{H.~P.}}
(\byear{2017}).
\btitle{Optimal sequential detection in multi-stream data}.
\bjournal{The Annals of Statistics}
\bvolume{45}
\bpages{2736--2763}.
\end{barticle}
\endbibitem

\bibitem{chan2015optimal}
\begin{barticle}[author]
\bauthor{\bsnm{Chan},~\bfnm{Hock~Peng}\binits{H.~P.}} \AND
  \bauthor{\bsnm{Walther},~\bfnm{Guenther}\binits{G.}}
(\byear{2015}).
\btitle{{Optimal detection of multi-sample aligned sparse signals}}.
\bjournal{The Annals of Statistics}
\bvolume{43}
\bpages{1865 -- 1895}.
\bdoi{10.1214/15-AOS1328}
\end{barticle}
\endbibitem

\bibitem{chen2022high}
\begin{barticle}[author]
\bauthor{\bsnm{Chen},~\bfnm{Yudong}\binits{Y.}},
  \bauthor{\bsnm{Wang},~\bfnm{Tengyao}\binits{T.}} \AND
  \bauthor{\bsnm{Samworth},~\bfnm{Richard~J}\binits{R.~J.}}
(\byear{2022}).
\btitle{High-dimensional, multiscale online changepoint detection}.
\bjournal{Journal of the Royal Statistical Society Series B: Statistical
  Methodology}
\bvolume{84}
\bpages{234--266}.
\end{barticle}
\endbibitem

\bibitem{delaigle2011robustness}
\begin{barticle}[author]
\bauthor{\bsnm{Delaigle},~\bfnm{Aurore}\binits{A.}},
  \bauthor{\bsnm{Hall},~\bfnm{Peter}\binits{P.}} \AND
  \bauthor{\bsnm{Jin},~\bfnm{Jiashun}\binits{J.}}
(\byear{2011}).
\btitle{Robustness and accuracy of methods for high dimensional data analysis
  based on {S}tudent’s t-Statistic}.
\bjournal{Journal of the Royal Statistical Society Series B: Statistical
  Methodology}
\bvolume{73}
\bpages{283-301}.
\bdoi{10.1111/j.1467-9868.2010.00761.x}
\end{barticle}
\endbibitem

\bibitem{donoho2004higher}
\begin{barticle}[author]
\bauthor{\bsnm{Donoho},~\bfnm{David}\binits{D.}} \AND
  \bauthor{\bsnm{Jin},~\bfnm{Jiashun}\binits{J.}}
(\byear{2004}).
\btitle{{Higher criticism for detecting sparse heterogeneous mixtures}}.
\bjournal{The Annals of Statistics}
\bvolume{32}
\bpages{962 -- 994}.
\bdoi{10.1214/009053604000000265}
\end{barticle}
\endbibitem

\bibitem{donoho2008higher}
\begin{barticle}[author]
\bauthor{\bsnm{Donoho},~\bfnm{David}\binits{D.}} \AND
  \bauthor{\bsnm{Jin},~\bfnm{Jiashun}\binits{J.}}
(\byear{2008}).
\btitle{Higher criticism thresholding: Optimal feature selection when useful
  features are rare and weak}.
\bjournal{Proceedings of the National Academy of Sciences}
\bvolume{105}
\bpages{14790--14795}.
\end{barticle}
\endbibitem

\bibitem{donoho2009feature}
\begin{barticle}[author]
\bauthor{\bsnm{Donoho},~\bfnm{David}\binits{D.}} \AND
  \bauthor{\bsnm{Jin},~\bfnm{Jiashun}\binits{J.}}
(\byear{2009}).
\btitle{Feature selection by higher criticism thresholding achieves the optimal
  phase diagram}.
\bjournal{Philosophical Transactions of the Royal Society A: Mathematical,
  Physical and Engineering Sciences}
\bvolume{367}
\bpages{4449--4470}.
\end{barticle}
\endbibitem

\bibitem{donoho2015higher}
\begin{barticle}[author]
\bauthor{\bsnm{Donoho},~\bfnm{David}\binits{D.}} \AND
  \bauthor{\bsnm{Jin},~\bfnm{Jiashun}\binits{J.}}
(\byear{2015}).
\btitle{{Higher criticism for large-scale inference, especially for rare and
  weak effects}}.
\bjournal{Statistical Science}
\bvolume{30}
\bpages{1 -- 25}.
\bdoi{10.1214/14-STS506}
\end{barticle}
\endbibitem

\bibitem{DonohoKipnis2020}
\begin{barticle}[author]
\bauthor{\bsnm{Donoho},~\bfnm{David}\binits{D.}} \AND
  \bauthor{\bsnm{Kipnis},~\bfnm{Alon}\binits{A.}}
(\byear{2022}).
\btitle{Higher criticism to compare two large frequency tables, with
  sensitivity to possible rare and weak differences}.
\bjournal{The Annals of Statistics}
\bvolume{50}
\bpages{1447--1472}.
\end{barticle}
\endbibitem

\bibitem{gosmann2022sequential}
\begin{barticle}[author]
\bauthor{\bsnm{G{\"o}smann},~\bfnm{Josua}\binits{J.}},
  \bauthor{\bsnm{Stoehr},~\bfnm{Christina}\binits{C.}},
  \bauthor{\bsnm{Heiny},~\bfnm{Johannes}\binits{J.}} \AND
  \bauthor{\bsnm{Dette},~\bfnm{Holger}\binits{H.}}
(\byear{2022}).
\btitle{Sequential change point detection in high dimensional time series}.
\bjournal{Electronic Journal of Statistics}
\bvolume{16}
\bpages{3608--3671}.
\end{barticle}
\endbibitem

\bibitem{hall2008properties}
\begin{barticle}[author]
\bauthor{\bsnm{Hall},~\bfnm{Peter}\binits{P.}} \AND
  \bauthor{\bsnm{Jin},~\bfnm{Jiashun}\binits{J.}}
(\byear{2008}).
\btitle{Properties of higher criticism under strong dependence}.
\bjournal{The Annals of Statistics}
\bvolume{36}
\bpages{381--402}.
\end{barticle}
\endbibitem

\bibitem{hall2010innovated}
\begin{barticle}[author]
\bauthor{\bsnm{Hall},~\bfnm{Peter}\binits{P.}} \AND
  \bauthor{\bsnm{Jin},~\bfnm{Jiashun}\binits{J.}}
(\byear{2010}).
\btitle{Innovated higher criticism for detecting sparse signals in correlated
  noise}.
\bjournal{The Annals of Statistics}
\bvolume{38}
\bpages{1686--1732}.
\end{barticle}
\endbibitem

\bibitem{hu2023likelihood}
\begin{barticle}[author]
\bauthor{\bsnm{Hu},~\bfnm{Shouri}\binits{S.}},
  \bauthor{\bsnm{Huang},~\bfnm{Jingyan}\binits{J.}},
  \bauthor{\bsnm{Chen},~\bfnm{Hao}\binits{H.}} \AND
  \bauthor{\bsnm{Chan},~\bfnm{Hock~Peng}\binits{H.~P.}}
(\byear{2023}).
\btitle{Likelihood scores for sparse signal and change-point detection}.
\bjournal{IEEE Transactions on Information Theory}
\bvolume{69}
\bpages{4065--4080}.
\end{barticle}
\endbibitem

\bibitem{jin2016rare}
\begin{barticle}[author]
\bauthor{\bsnm{Jin},~\bfnm{Jiashun}\binits{J.}} \AND
  \bauthor{\bsnm{Ke},~\bfnm{Zheng~Tracy}\binits{Z.~T.}}
(\byear{2016}).
\btitle{Rare and weak effects in large-scale inference: Methods and phase
  diagrams}.
\bjournal{Statistica Sinica}
\bpages{1--34}.
\end{barticle}
\endbibitem

\bibitem{kipnis2022higher}
\begin{barticle}[author]
\bauthor{\bsnm{Kipnis},~\bfnm{Alon}\binits{A.}}
(\byear{2022}).
\btitle{Higher criticism for discriminating word-frequency tables and
  authorship attribution}.
\bjournal{The Annals of Applied Statistics}
\bvolume{16}
\bpages{1236--1252}.
\end{barticle}
\endbibitem

\bibitem{kipnis2021logchisquared}
\begin{barticle}[author]
\bauthor{\bsnm{Kipnis},~\bfnm{Alon}\binits{A.}}
(\byear{2023}).
\btitle{Unification of rare/weak detection models using moderate deviations
  analysis and log-Chisquared P-values}.
\bjournal{Statistica Scinica}.
\bnote{To appear}.
\bdoi{10.5705/ss.202023.0128}
\end{barticle}
\endbibitem

\bibitem{lai1998information}
\begin{barticle}[author]
\bauthor{\bsnm{Lai},~\bfnm{Tze~Leung}\binits{T.~L.}}
(\byear{1998}).
\btitle{Information bounds and quick detection of parameter changes in
  stochastic systems}.
\bjournal{IEEE Transactions on Information Theory}
\bvolume{44}
\bpages{2917--2929}.
\end{barticle}
\endbibitem

\bibitem{li2015higher}
\begin{barticle}[author]
\bauthor{\bsnm{Li},~\bfnm{Jian}\binits{J.}} \AND
  \bauthor{\bsnm{Siegmund},~\bfnm{David}\binits{D.}}
(\byear{2015}).
\btitle{Higher criticism: $p$-values and criticism}.
\bjournal{The Annals of Statistics}
\bvolume{43}
\bpages{1323--1350}.
\end{barticle}
\endbibitem

\bibitem{liu2021minimax}
\begin{barticle}[author]
\bauthor{\bsnm{Liu},~\bfnm{Haoyang}\binits{H.}},
  \bauthor{\bsnm{Gao},~\bfnm{Chao}\binits{C.}} \AND
  \bauthor{\bsnm{Samworth},~\bfnm{Richard~J}\binits{R.~J.}}
(\byear{2021}).
\btitle{Minimax rates in sparse, high-dimensional change point detection}.
\bjournal{The Annals of Statistics}
\bvolume{49}.
\end{barticle}
\endbibitem

\bibitem{liu2019scalable}
\begin{barticle}[author]
\bauthor{\bsnm{Liu},~\bfnm{Kun}\binits{K.}},
  \bauthor{\bsnm{Zhang},~\bfnm{Ruizhi}\binits{R.}} \AND
  \bauthor{\bsnm{Mei},~\bfnm{Yajun}\binits{Y.}}
(\byear{2019}).
\btitle{Scalable sum-shrinkage schemes for distributed monitoring large-scale
  data streams}.
\bjournal{Statistica Sinica}
\bvolume{29}
\bpages{1--22}.
\end{barticle}
\endbibitem

\bibitem{lorden1971procedures}
\begin{barticle}[author]
\bauthor{\bsnm{Lorden},~\bfnm{G}\binits{G.}}
(\byear{1971}).
\btitle{Procedures for reacting to a change in distribution}.
\bjournal{The Annals of Mathematical Statistics}
\bvolume{42}
\bpages{1897--1908}.
\end{barticle}
\endbibitem

\bibitem{mei2010efficient}
\begin{barticle}[author]
\bauthor{\bsnm{Mei},~\bfnm{Yajun}\binits{Y.}}
(\byear{2010}).
\btitle{Efficient scalable schemes for monitoring a large number of data
  streams}.
\bjournal{Biometrika}
\bvolume{97}
\bpages{419--433}.
\end{barticle}
\endbibitem

\bibitem{pilliat2023optimal}
\begin{barticle}[author]
\bauthor{\bsnm{Pilliat},~\bfnm{Emmanuel}\binits{E.}},
  \bauthor{\bsnm{Carpentier},~\bfnm{Alexandra}\binits{A.}} \AND
  \bauthor{\bsnm{Verzelen},~\bfnm{Nicolas}\binits{N.}}
(\byear{2023}).
\btitle{Optimal multiple change-point detection for high-dimensional data}.
\bjournal{Electronic Journal of Statistics}
\bvolume{17}
\bpages{1240--1315}.
\end{barticle}
\endbibitem

\bibitem{pollak1985optimal}
\begin{barticle}[author]
\bauthor{\bsnm{Pollak},~\bfnm{Moshe}\binits{M.}}
(\byear{1985}).
\btitle{Optimal detection of a change in distribution}.
\bjournal{The Annals of Statistics}
\bvolume{13}
\bpages{206--227}.
\end{barticle}
\endbibitem

\bibitem{poor1998quickest}
\begin{barticle}[author]
\bauthor{\bsnm{Poor},~\bfnm{H~Vincent}\binits{H.~V.}}
(\byear{1998}).
\btitle{Quickest detection with exponential penalty for delay}.
\bjournal{The Annals of Statistics}
\bvolume{26}
\bpages{2179--2205}.
\end{barticle}
\endbibitem

\bibitem{shorack2009empirical}
\begin{bbook}[author]
\bauthor{\bsnm{Shorack},~\bfnm{Galen~R}\binits{G.~R.}} \AND
  \bauthor{\bsnm{Wellner},~\bfnm{Jon~A}\binits{J.~A.}}
(\byear{2009}).
\btitle{Empirical processes with applications to statistics}.
\bpublisher{SIAM}.
\end{bbook}
\endbibitem

\bibitem{siegmund1995using}
\begin{barticle}[author]
\bauthor{\bsnm{Siegmund},~\bfnm{David}\binits{D.}} \AND
  \bauthor{\bsnm{Venkatraman},~\bfnm{ES}\binits{E.}}
(\byear{1995}).
\btitle{Using the generalized likelihood ratio statistic for sequential
  detection of a change-point}.
\bjournal{The Annals of Statistics}
\bpages{255--271}.
\end{barticle}
\endbibitem

\bibitem{stoepker2023sparse}
\begin{barticle}[author]
\bauthor{\bsnm{Stoepker},~\bfnm{Ivo~V}\binits{I.~V.}},
  \bauthor{\bsnm{Castro},~\bfnm{Rui~M}\binits{R.~M.}} \AND
  \bauthor{\bsnm{Arias-Castro},~\bfnm{Ery}\binits{E.}}
(\byear{2025}).
\btitle{Sparse dnomaly detection across referentials: A rank-based higher
  criticism approach}.
\bjournal{The Annals of Statistics}.
\bnote{To appear}.
\end{barticle}
\endbibitem

\bibitem{stoepker2024anomaly}
\begin{barticle}[author]
\bauthor{\bsnm{Stoepker},~\bfnm{Ivo~V}\binits{I.~V.}},
  \bauthor{\bsnm{Castro},~\bfnm{Rui~M}\binits{R.~M.}},
  \bauthor{\bsnm{Arias-Castro},~\bfnm{Ery}\binits{E.}} \AND
  \bauthor{\bparticle{van~den} \bsnm{Heuvel},~\bfnm{Edwin}\binits{E.}}
(\byear{2024}).
\btitle{Anomaly detection for a large number of streams: A permutation-based
  higher criticism approach}.
\bjournal{Journal of the American Statistical Association}
\bvolume{119}
\bpages{461--474}.
\end{barticle}
\endbibitem

\bibitem{tartakovsky2006detection}
\begin{barticle}[author]
\bauthor{\bsnm{Tartakovsky},~\bfnm{Alexander~G}\binits{A.~G.}},
  \bauthor{\bsnm{Rozovskii},~\bfnm{Boris~L}\binits{B.~L.}},
  \bauthor{\bsnm{Bla{\v{z}}ek},~\bfnm{Rudolf~B}\binits{R.~B.}} \AND
  \bauthor{\bsnm{Kim},~\bfnm{Hongjoong}\binits{H.}}
(\byear{2006}).
\btitle{Detection of intrusions in information systems by sequential
  change-point methods}.
\bjournal{Statistical Methodology}
\bvolume{3}
\bpages{252--293}.
\end{barticle}
\endbibitem

\bibitem{tartakovsky2005general}
\begin{barticle}[author]
\bauthor{\bsnm{Tartakovsky},~\bfnm{Alexander~G}\binits{A.~G.}} \AND
  \bauthor{\bsnm{Veeravalli},~\bfnm{Venugopal~V}\binits{V.~V.}}
(\byear{2005}).
\btitle{General asymptotic Bayesian theory of quickest change detection}.
\bjournal{Theory of Probability \& Its Applications}
\bvolume{49}
\bpages{458--497}.
\end{barticle}
\endbibitem

\bibitem{tartakovsky2008asymptotically}
\begin{barticle}[author]
\bauthor{\bsnm{Tartakovsky},~\bfnm{Alexander~G}\binits{A.~G.}} \AND
  \bauthor{\bsnm{Veeravalli},~\bfnm{Venugopal~V}\binits{V.~V.}}
(\byear{2008}).
\btitle{Asymptotically optimal quickest change detection in distributed sensor
  systems}.
\bjournal{Sequential Analysis}
\bvolume{27}
\bpages{441--475}.
\end{barticle}
\endbibitem

\bibitem{tony2011optimal}
\begin{barticle}[author]
\bauthor{\bsnm{Tony~Cai},~\bfnm{T.}\binits{T.}},
  \bauthor{\bsnm{Jessie~Jeng},~\bfnm{X.}\binits{X.}} \AND
  \bauthor{\bsnm{Jin},~\bfnm{Jiashun}\binits{J.}}
(\byear{2011}).
\btitle{Optimal detection of heterogeneous and heteroscedastic mixtures}.
\bjournal{Journal of the Royal Statistical Society Series B: Statistical
  Methodology}
\bvolume{73}
\bpages{629-662}.
\bdoi{10.1111/j.1467-9868.2011.00778.x}
\end{barticle}
\endbibitem

\bibitem{xie2021sequential}
\begin{barticle}[author]
\bauthor{\bsnm{Xie},~\bfnm{Liyan}\binits{L.}},
  \bauthor{\bsnm{Zou},~\bfnm{Shaofeng}\binits{S.}},
  \bauthor{\bsnm{Xie},~\bfnm{Yao}\binits{Y.}} \AND
  \bauthor{\bsnm{Veeravalli},~\bfnm{Venugopal~V}\binits{V.~V.}}
(\byear{2021}).
\btitle{Sequential (quickest) change detection: Classical results and new
  directions}.
\bjournal{IEEE Journal on Selected Areas in Information Theory}
\bvolume{2}
\bpages{494--514}.
\end{barticle}
\endbibitem

\bibitem{xie2013sequential}
\begin{barticle}[author]
\bauthor{\bsnm{Xie},~\bfnm{Yao}\binits{Y.}} \AND
  \bauthor{\bsnm{Siegmund},~\bfnm{David}\binits{D.}}
(\byear{2013}).
\btitle{{Sequential multi-sensor change-point detection}}.
\bjournal{The Annals of Statistics}
\bvolume{41}
\bpages{670 -- 692}.
\bdoi{10.1214/13-AOS1094}
\end{barticle}
\endbibitem

\bibitem{xu2022active}
\begin{binproceedings}[author]
\bauthor{\bsnm{Xu},~\bfnm{Qunzhi}\binits{Q.}} \AND
  \bauthor{\bsnm{Mei},~\bfnm{Yajun}\binits{Y.}}
(\byear{2022}).
\btitle{Active quickest detection when monitoring multi-streams with two
  affected streams}.
In \bbooktitle{2022 IEEE International Symposium on Information Theory (ISIT)}
\bpages{1915--1920}.
\bpublisher{IEEE}.
\end{binproceedings}
\endbibitem

\bibitem{yakir2013extremes}
\begin{bbook}[author]
\bauthor{\bsnm{Yakir},~\bfnm{Benjamin}\binits{B.}}
(\byear{2013}).
\btitle{Extremes in random fields: A theory and its applications}.
\bpublisher{John Wiley \& Sons}.
\end{bbook}
\endbibitem

\end{thebibliography}



\clearpage
\begin{appendix}
\section{Proofs}
\label{sec:proofs}

\subsection{Technical Lemmas}
We provide here several technical lemmas that will be useful in the proof of Theorem~\ref{thm:alternative}. Lemmas~\ref{lem:argmax_small}-\ref{lem:argmax_GLR} focus on univariate sequential change-point detection using the LR and GLR statistics of \eqref{eq:YLR_def} and \eqref{eq:YGLR_def}. For convenience, we assume in these lemmas that $n=1$.

The following lemma shows that the maximum in $Y_t^{\GLR}$ and $Y_t^{\LR}$ is attained after the change with probability approaching one, provided $t$ does not grow faster than the increase in the mean shift. This property seems well-known for $Y_t^{\LR}$ based on the discussion in \cite{siegmund1995using}. The proof is somewhat simpler than in previous cases because the individual effect size $\mu_N$ goes to infinity in $N$. This lemma justifies the analysis of the P-values under $H_1$. 

\begin{lemma}
\label{lem:argmax_small}
Consider $V_{1,t,k}$ of \eqref{eq:YLR_def}
and $W_{1,t,k}$ of \eqref{eq:YGLR_def}. 
Suppose that $X_{1,1},\ldots,X_{1,\tau-1} \simiid \Ncal(0,1)$ and $X_{1,\tau},\ldots,X_{1,\mathcal T} \simiid \Ncal(\mu, \sigma^2)$ for some $1 \leq \tau \leq \Tcal$, $\sigma^2>0$. Assume that $\mu \to \infty$ and $\Tcal \to \infty$ while
$\sqrt{\log(\Tcal)}/\mu \to 0$. Then, 
\[
    \Prp{\arg \max_{k \leq t} V_{1,t,k} < \tau } = o(1)
\]
and 
    \[
    \Prp{\arg \max_{k \leq t} 
    W_{1,t,k}
    < \tau } = o(1). 
    \]
\end{lemma}
(here and throughout this section, 
$o(1)$ represents an expression converging to $0$ uniformly in $t,\tau \leq \Tcal$ as $\mu \to \infty$.)  

\begin{proof}[Proof of Lemma \ref{lem:argmax_small}]
Denote $X_n := X_{1,n}$ and consider the following events defined by $X_1,\ldots,X_{\Tcal}$.
\begin{align*}
    \Omega_{V_t} &:= \{ \arg \max_{k \leq t} V_{1,t,k} < \tau\},\\
    \Omega_{W_t} &:= \{ \arg \max_{k \leq t} W_{1,t,k} < \tau\}, \\
    B_t & := \left\{ \text{$\exists$ $0<m<\tau$, $0\leq k\leq t-\tau$, such that $X_{\tau - m} \geq X_{\tau + k}$ } \right\}. 
\end{align*}
Note that $\Omega_{V_t} \subset B_t$ and $\Omega_{W_t} \subset B_t$. Consequently, it is enough to show that $\Prp{B_t} \to 0$. 

Let $Z_1,\ldots,Z_m \simiid \Ncal(\mu,\sigma^2)$. Then
\begin{align*}
\Prp{\min\{Z_1,\ldots,Z_m\} \leq a} & = 1 - \left(1 - \Phi\left(\frac{a-\mu}{\sigma}\right) \right)^m = 1 - \Phi^m\left(\frac{\mu-a}{\sigma}\right). \\
\Prp{\max\{Z_1,\ldots,Z_m\} \leq a} & = \Phi^m\left(\frac{a-\mu}{\sigma}\right).
\end{align*}
Define $A_a = \{ \max\{X_1, \ldots X_{\tau-1}\} \leq a\}$. 
We have 
    \begin{align*}
        \Prp{B_t} & = \Prp{ \max\{X_1,\ldots,X_{\tau-1}\} \geq \min\{X_{\tau},\ldots,X_{\Tcal}\}} \\
        & \leq \Prp{ \max\{X_1,\ldots,X_{\tau-1}\} \geq \min\{X_{\tau},\ldots,X_{\Tcal}\} \mid A_a } \Prp{A_a} + \Prp{A_a^c} \\
        & \leq  \Prp{ a \geq \min\{X_{\tau},\ldots,X_{\Tcal}\} \mid A_a } + \Prp{A_a^c} \\
        & = 1 - \Phi^{\Tcal - \tau+1}\left( \frac{\mu-a}{\sigma} \right) + \Prp{A_a^c}. 
    \end{align*}
    Let $a = \sqrt{4 \log \Tcal}$, then 
    recall that $\Phi(x)\sim1-\phi(x)/x$
    \[
    \Prp{A_a} = \Phi^{\tau-1}(\sqrt{4 \log \Tcal}) \geq \Phi^\Tcal(\sqrt{4 \log \Tcal})
    \sim \left(1-\frac{e^{-2\log\Tcal}}{\sqrt{8\pi\log\Tcal}}\right)^{\Tcal} = 1 + o(1), 
    \]
    hence $\Prp{A_a^c} \to 0$ as $ N \to \infty$. Additionally, since $\mu \to \infty$ and $\mu/\sqrt{\log(\Tcal)} \to \infty$, also $\mu - \sqrt{4\log(\Tcal)} \to \infty$ and we can use the upper bound $1 - \Phi(x) \leq e^{-x^2/2}/\sqrt{2\pi}$ as $x \to \infty$. This implies that for all $N$ large enough, 
    \begin{align*}
    \Phi^{\Tcal - \tau}\left( \frac{\mu_N-\sqrt{4 \log \Tcal}}{\sigma} \right)
         & \geq \Phi^\Tcal\left( \frac{\mu_N-\sqrt{4 \log \Tcal}}{\sigma} \right) 
         \\
         & \sim \left( 1- e^{-\frac{1}{2}\left(\mu_N - \sqrt{4 \log(\Tcal)}\right)^2}\right)^\Tcal\\
         & = \left( 1- e^{-2 \log(\Tcal) \left(\frac{\mu_N}{\sqrt{4 \log(\Tcal)}} - 1\right)^2 } \right)^\Tcal \\
         & \geq  \left( 1- e^{-2 \log(\Tcal)} \right)^\Tcal = 1 + o(1),
    \end{align*}
    where the last transition is because $\mu/\sqrt{\log(\Tcal)} \to \infty$. We conclude that $\Prp{B_t} = o(1)$ which completes the proof of the lemma. 
\end{proof}

\begin{lemma}
\label{lem:argmax_large_V}
Consider the setting and assumptions of Lemma~\ref{lem:argmax_small}. Then 
\[
    \lim_{\mu \to \infty} \Prp{ Y_{1,t}^{\LR} = \max_{\tau \leq k \leq t} V_{1,t,k} } = 1 + o(1),
\]
uniformly in $\tau,t \leq \Tcal$. 
\end{lemma}

\begin{proof}[Proof of Lemma \ref{lem:argmax_large_V}]
Denote $W_{t,k} := W_{1,t,k}$, $Y_{t}^{\LR}:=Y_{1,t}^{\LR}$, $X_t := X_{1,t}$, and $S_{1,t} := S_t$.

Notice that $W_{t,m} =\mu\left[\sum_{j=m+1}^t X_j-\frac{\mu}{2}(t-m)\right]$ is distributed as $\mathcal N(\frac{\mu^2}{2}(t-m),\mu^2(t-m))$ under $H_1$ for $\tau \leq m \leq t$. For $\tau \leq m \leq t$, 
\begin{align}
    \Prp{W_{t,m} < 0} &= \Prp{ \Ncal\left(\mu^2(t-m)/2,\mu^2 (t-m) \right) < 0 } \nonumber  \\
    & = \Phi\left(- \mu\sqrt{t-m}/2\right) = e^{-\mu^2(t-m)/8}(1+o(1)). 
    \label{eq:argmax_large_V:proof:1}
\end{align}
In addition,
\begin{align*}
    \max_{k \leq m} V_{m,k} - \max_{k \leq t} V_{t,k} & \leq 
    \max_{k \leq m} (V_{m,k} - V_{t,k}) = -V_{t,m}.
\end{align*}
Therefore, 
\[
\Prp{
\max_{k \leq m} V_{m,k} - \max_{k \leq t} V_{t,k} >0} \leq \Prp{-V_{t,m}>0} =
e^{-\mu^2(t-m)/8}(1+o(1)). 
\]
where the last equality is due to \eqref{eq:argmax_large_V:proof:1}.
Consider the event
\begin{align*}
    B_{u,t} := \{ \exists\, u \leq m < t\,:\, \max_{k\leq m} V_{m,k} > \max_{k\leq t} V_{t,k} \}. 
\end{align*}
Notice that
\[
\Prp{Y_t^{\LR} > \max_{\tau \leq k \leq t} V_{t,k}} \leq \Prp{B_{1,t}} \leq \Prp{B_{1,\tau}} + \Prp{B_{\tau,t}}. 
\]
The first inequality is because  
\begin{align*}
\{
Y_t^{\LR} > \max_{\tau \leq k \leq t} V_{t,k}
\} & = \{
\max_{1 \leq k \leq t} V_{t,k} > \max_{\tau \leq k \leq t} V_{t,k}
\} \\
& = 
\{
\max_{1 \leq k \leq \tau} V_{t,k} > \max_{\tau \leq k \leq t} V_{t,k}
\} \\
& =
\{
\max_{1 \leq k \leq \tau} V_{t,k} \geq \max_{1 \leq k \leq t} V_{t,k}
\},
\end{align*}
and the last event by definition is $B_{1,t}$. We have $\Prp{B_{1,\tau}} = o(1)$ by similar arguments as in Lemma~\ref{lem:argmax_small}. Additionally, 
\begin{align*}
\Prp{B_{\tau,t}} & \leq o(1) + \sum_{m=\tau}^{t-1}\Prp{ \max_{k \leq m} V_{m,k} - \max_{k \leq t} V_{t,k} >0} \\
& \leq (1+o(1))\sum_{m=\tau}^{t-1} e^{-\mu^2(t-m)/8} \\
& \leq (1+o(1)) \frac{e^{-\mu^2/8}}{1 - e^{-\mu^2/8}} = o(1).
\end{align*}
By definition, we have that $Y_t^{\LR} \geq \max_{\tau \leq k \leq t} V_{t,k}$. We conclude 
\[
\Prp{Y_t^{\LR} \neq \max_{\tau \leq k \leq t} V_{t,k} } = \Prp{Y_t^{\LR} > \max_{\tau \leq k \leq t} V_{t,k} } = o(1).
\]
This completes the proof.
\end{proof}

The following claim says that with probability going to $1$, the maximum in the LR statistic is attained at the change point $k=\tau$. 
\begin{lemma} 
\label{lem:argmax_LR}
Consider the setting and assumption of Lemma~\ref{lem:argmax_large_V}. Then, 
    \begin{align*}
         \Prp{ Y_{1,t}^{\LR} = V_{1,t,\tau} }  = 1 + o(1).
    \end{align*}
\end{lemma}

\begin{proof}[Proof of Lemma \ref{lem:argmax_LR}]
Denote $V_{t,k} := V_{1,t,k}$, $Y_{t}^{\LR}:=Y_{1,t}^{\LR}$, and $S_{1,t} := S_t$. By Lemma~\ref{lem:argmax_large_V}, we may assume 
    \[
    Y_{t}^{\LR} = \max_{\tau \leq k \leq t} V_{t,k},
    \]
    almost surely, since the probability of the complementary event vanishes as $\mu \to \infty$.
Since $Y_t^{\LR} \geq V_{t,\tau}$, it is sufficient to show that $\Prp{Y_t^{\LR} \geq V_{t,k} + \epsilon} \to 0$ for any $\epsilon>0$. In the following, $Z_1,\ldots Z_m$ is a sequence of independent standard normal random variables. We also use the notational convention $\sum_{i=1}^0 Z_i \equiv 0$ and assume without loss of generality that $\mu>0$. Fix $\epsilon>0$. 
    \begin{align*}
      \Prp{ Y_t^{\LR} \geq V_{t,\tau} + \epsilon } &  = \Prp{ \max_{\tau \leq k \leq t} \left\{{S_t - S_k - \frac{\mu}{2}(t-k)} \right\}\geq S_t - S_\tau - \frac{\mu}{2}(t-\tau) + \frac{\epsilon}{\mu} } \\
      & = \Prp{ \max_{\tau \leq k \leq t}{\left\{ {\color{black} S_\tau - S_k - \frac{\mu}{2}(\tau-k)}\right\}} \geq \frac{\epsilon}{\mu}} \\
      & = \Prp{ \max_{\tau \leq k \leq t}\left\{ -\left(S_k - S_{\tau} - \frac{\mu}{2}(k - \tau) \right) \right\} \geq \frac{\epsilon}{\mu}} \\
      & = \Prp{ \max_{\tau \leq k \leq t}\left\{ -\left(S_k - S_{\tau} - \mu(k - \tau) \right) - \frac{\mu}{2}(k - \tau) \right\} \geq \frac{\epsilon}{\mu}} \\
      & = \Prp{ \min_{0\leq m \leq t - \tau} \left\{ \sigma \sum_{i=1}^m Z_i + \frac{\mu}{2}m \right\} \leq - \frac{\epsilon}{\mu}} \\
      & = \Pr \left[ 
      \left\{ \sigma Z_1 + \frac{\mu}{2} \leq -\frac{\epsilon}{\mu} \right\} 
      \cup \left\{ \sigma(Z_1 + Z_2) + 2\frac{\mu}{2} \leq -\frac{\epsilon}{\mu} \right\} \cup \right. \cdots \\
      & \qquad \qquad \left. \cup \left\{ \sigma \sum_{i=1}^{t-\tau} Z_i + (t-\tau)\frac{\mu}{2} \leq -\frac{\epsilon}{\mu} \right\} \right]\\
      & \leq \sum_{m=1}^{t-\tau} \Prp{\sigma \sum_{i=1}^m Z_i + m \frac{\mu}{2} \leq -\frac{\epsilon}{\mu} } = \sum_{m=1}^{t-\tau} \Prp{\sigma\sqrt{m}Z_1 + m\frac{\mu}{2} \leq -\frac{\epsilon}{\mu} } \\
      & = \sum_{m=1}^{t-\tau} \Phi\left( \frac{-\frac{\epsilon}{\mu} - m \mu/2}{\sigma \sqrt{m}} \right) \leq \sum_{m=1}^{t-\tau} \Phi\left(- \frac{\sqrt{m} \mu}{2\sigma} \right) \\
      & \leq \frac{1}{\sigma\sqrt{2\pi}} \sum_{m=1}^{t-\tau} e^{-m \mu^2 /(8\sigma^2)} \leq \frac{1}{\sigma\sqrt{2\pi}} \frac{e^{-\mu^2 /(8\sigma^2)}}{1 - e^{-\mu^2 /(8\sigma^2)}}. 
    \end{align*}
    The proof is completed by noticing that, as $\mu\to\infty$, the last expression converges to zero and is independent of $t$.
\end{proof}

\begin{lemma}
\label{lem:bound_Y_t_v}
Consider $W_{1,t,k}$ of \eqref{eq:YGLR_def}. 
Suppose that $X_{1,\tau},\ldots,X_{1,\mathcal T} \simiid \Ncal(\mu, \sigma^2)$ for some $1 \leq \tau \leq \Tcal$, $\sigma^2>0$, and $\mu \to \infty$. For $\tau \leq v \leq t$
    \begin{align}
        \Prp{W_{1,t,\tau} \leq W_{1,t,v}} \leq \exp\left\{ -\frac{1}{2} \frac{\mu^2(v-\tau)}{5 \sigma^2} \right\}(1+o(1)).
        \label{eq:bound_Y_t_v}
    \end{align}
\end{lemma}

The following lemma bounds the probability of events of the form $ W_{1,t,\tau} < W_{1,t,v}$ for $v < t$ after the change as $\mu \to \infty$. The bound behaves as if $J_k := W_{1,t,k}$ is a Gaussian random walk in this limit. 
\begin{proof}[Proof of Lemma \ref{lem:bound_Y_t_v}]
Denote $W_{t,k} := W_{1,t,k}$, $Y_{t}^{\LR}:=Y_{1,t}^{\LR}$, $X_t := X_{1,t}$, and $S_{1,t} := S_t$. In the following, for integers $\tau < v<t$, denote $S_{t,v} := S_t - S_v$. We have
    \begin{align*}
    \Prp{ W_{t,\tau} < W_{t,v}} & = \Prp{ \frac{S^2_{t,\tau}}{t-\tau} < \frac{S^2_{t,v}}{t-v} } \\
    & = \Prp{ \frac{S^2_{t,v}}{t-\tau} + \frac{S^2_{v,\tau}}{t-\tau} + 2 \frac{S_{t,v} S_{v,\tau}}{t-\tau}  < \frac{S^2_{t,v}}{t-v} } \\
    & = \Prp{ S^2_{t,v} \frac{v-\tau}{t-v} >  S^2_{v,\tau}+ 2 S_{t,v} S_{v,\tau} } 
    \end{align*}
    where the second equality uses the fact that $S_{t, \tau} = S_{t, v} + S_{v, \tau}$ and the third equality follows by grouping terms with $S_{t,v}^2$. 
    We now solve the quadratic equation for $S_{t,v}$. 
    This leads to 
    \begin{align*}
    \Prp{ W_{t,\tau} < W_{t,v}} 
    & \leq \Prp{ S_{t,v}\frac{v-\tau}{t-v} > S_{v,\tau}\left(1 + \sqrt{ 1 + \frac{v-\tau}{t-v} } \right) } \\
    & \qquad + \Prp{ S_{t,v} \frac{v-\tau}{t-v}< S_{v,\tau}\left(1 - \sqrt{ 1 + \frac{v-\tau}{t-v} } \right)}.
    \end{align*}
    The second term goes to zero as $N$ goes to infinity whenever $v > \tau$ because the probability that one of $S_{t,v}$ and $S_{v,\tau}$ is negative goes to zero as their mean increases to infinity. We proceed by handling the first term. We denote by $Z_1$, $Z_2$, and $Z_3$ three independent standard normal RVs. Using that $S_{v,\tau}$ is independent of $S_{t,v}$ and both are sums of several independent $\Ncal(\mu, \sigma^2)$, we get
    \begin{align*}
        & \Prp{ S_{t,v} \frac{v-\tau}{t-v} > S_{v,\tau} \left(1 + \sqrt{ 1 + \frac{v-\tau}{t-v} } \right)} \\
        & = \Prp{ \frac{S_{t,v}}{t-v} >  \frac{S_{v,\tau}}{v-\tau} \left(1 + \sqrt{ 1 + \frac{v-\tau}{t-v} } \right) } \\
        & = \Prp{  \frac{ \sigma Z_1}{\sqrt{t-v}} + \mu > ( \frac{ \sigma Z_2}{\sqrt{v-\tau}} + \mu) \left(1 + \sqrt{ 1 + \frac{v-\tau}{t-v} } \right)  } \\ 
        & = \Prp{ \frac{ Z_1}{\sqrt{t-v}} > \frac{Z_2}{\sqrt{v-\tau}} \left(1 + \sqrt{ 1 + \frac{v-\tau}{t-v} } \right) + \frac{\mu}{\sigma}\sqrt{1 + \frac{v-\tau}{t-v}} } \\
        & = \Prp{ Z_1 \sqrt{v-\tau} > Z_2 \left(\sqrt{t-v} + \sqrt{t-\tau} \right) + \frac{\mu}{\sigma}\sqrt{t-\tau} \sqrt{v-\tau} } \\
        & = \Prp{ Z_3 \geq \frac{\mu}{\sigma}\sqrt{
        \frac{(t-\tau)(v-\tau)}{v-\tau + \left( \sqrt{t-v} + \sqrt{t-\tau} \right)^2}
        }} 
         \\
        & \leq \Prp{ Z_3 \geq \frac{\mu}{\sigma}\sqrt{
        \frac{(t-\tau)(v-\tau)}{t-\tau + \left( \sqrt{t-\tau} + \sqrt{t-\tau} \right)^2}
        }} \\
        & = \Prp{ Z_3 \geq \frac{\mu}{\sigma}\sqrt{\frac{v-\tau}{5}} 
        } = 1 - \Phi\left( \frac{\mu\sqrt{v-\tau}}{\sqrt{5}\sigma} \right). 
    \end{align*}
    From here, the advertised claim follows from Mill's ratio. 
\end{proof}

\begin{lemma}
\label{lem:argmax_large_Y}
Consider the setting and assumptions of Lemma~\ref{lem:argmax_small}. Then, 
\[
    \Prp{ Y_{1,t}^{\GLR} = \max_{\tau \leq k \leq t} W_{1,t,k} } = 1 + o(1).
\]
\end{lemma}

\begin{proof}[Proof of Lemma \ref{lem:argmax_large_Y}]
The proof is similar to Lemma~\ref{lem:argmax_large_V}. Simply replace \eqref{eq:argmax_large_V:proof:1} with 
\eqref{eq:bound_Y_t_v}. 
\end{proof}

The following claim says that with probability going to $1$, the maximum in the GLR statistic is attained at the change point $k=\tau$. 
\begin{lemma}
\label{lem:argmax_GLR}
Consider the setting and conditions of Lemma~\ref{lem:argmax_large_Y}. Then
    \begin{align*}
\Prp{ Y_{1,t}^{\GLR} = W_{1,t,\tau}} = 1 + o(1).
    \end{align*}
\end{lemma}

\begin{proof}[Proof of Lemma \ref{lem:argmax_GLR}]
Denote $W_{t,k} := W_{1,t,k}$ and  $Y_{t}^{\LR}:=Y_{1,t}^{\LR}$.
By Lemma~\ref{lem:argmax_large_Y}, we may assume
\[
Y_t^{\GLR} = \max_{\tau \leq k \leq t} W_{t,k},
\]
since the probability of the complementary event vanishes as $N \to \infty$.
For $\tau < v \leq t$, we use  Lemma~\ref{lem:bound_Y_t_v} and the union bound to conclude
    \begin{align*}
        & \Prp{ \exists v>\tau\,:\, W_{t,\tau} < W_{t,v}} \leq \Prp{ \{W_{t,\tau} < W_{t,\tau+1} \} \cup \cdots \{W_{t,\tau} < W_{t,t-1} \} } \\
        & \leq \sum_{v=\tau+1}^{t-1} \Prp{W_{t,\tau} < W_{t,v}} \leq \frac{1}{\sqrt{10 \pi \sigma^2 } } 
        \sum_{v=\tau+1}^{t-1} e^{-\frac{ \mu^2 (v - \tau)}{5\sigma^2}} \\
        & \leq \frac{1}{\sqrt{10 \pi \sigma^2 } } 
        \frac{e^{-\frac{ \mu^2 }{5\sigma^2}}}{1- e^{-\frac{ \mu^2/2}{5\sigma^2}}} \to 0, 
    \end{align*}
    uniformly in $t$ as $\mu \to \infty$.
\end{proof}

\subsection{Proof of Main Results}
\label{sec:proof_main}
\begin{proof}[Proof of Corollary \ref{cor:impossible_general}]
Denote by $\{\tilde{Q}_n^{(N)}\, n=1,\ldots,N\}$ a sequence of distributions that satisfies \eqref{eq:logchisqaured_appx} with $\rho = \rho^-$. Let $\tilde{\pi}_1,\ldots, \tilde{\pi}_N$ be P-values obeying \eqref{eq:hyp_test_pvalues} with $Q_{n}^{(N)} = \tilde{Q}_n^{(N)}$ and let $\tilde{U}^{(N)} = U(\tilde{\pi}_1,\ldots,\tilde{\pi}_{N})$ be some detection statistic based on these P-values.  

By \eqref{eq:p_value_rare_model} and $\rho(\Delta) \leq \rho^-$, for $t \in \Theta_0$ and $q > \rho^-$ we have the asymptotic inequality
\begin{align*}
    \Prp{ -2 \log(\tilde{Q}_n^{(N)}) \geq 2q\log(N) } = N^{-\left(\frac{\sqrt{q}-\sqrt{\rho^-}}{\sigma} \right)^2+o(1)} \geq \Prp{ -2 \log(Q_{n,t}^{(N)}) \geq 2q\log(N)}.
\end{align*}
Namely, every $-2 \log(Q_{n,t}^{(N)})$ is asymptotically stochastically bounded by $-2 \log(\tilde{Q}_n^{(N)})$. Thus, by \eqref{eq:p_value_rare_model_pval}, every $\pi_{n,t}$ is 
asymptotically stochastically bounded below by $\tilde{\pi}_n$. The assumption that $U$ is non-decreasing as its arguments decrease implies that $\tilde{U}^{(N)}$ stochastically dominates $U_t^{(N)}$ for all $t \leq \Tcal$ in the sense that for any $b>0$ and $N$, 
\begin{align}
    \label{eq:dominance}
\Prp{\tilde{U}^{(N)} \geq b \mid H_1^{(N)}} \geq \Prp{U_t^{(N)} \geq b \mid H_1^{(N)}}. 
\end{align}
Suppose by contradiction that there exists $a>0$ and $\{\Delta_k\}_{k=1,2,\ldots}$ such that
\begin{align}
\label{eq:proof:cor:impossible}
    \lim_{k \to \infty} \Prp{ T_U \leq \Delta_k \mid H_1^{(N_k)}} - \Prp{ T_U \leq \Delta_k \mid H_0^{(N_k)}} = a,
\end{align}
along some sub-sequence $N_1 < N_2 <...<N_k$, $k=1,2,\ldots$, where 
$T_U = \inf \{ t\,:\, U_t^{(N_k)} \geq b_t^{(N_k)} \}$ with $U_t^{(N_k)}$ based on $\{\pi_{n,t},\,n=1,\ldots,N_k \}$. Under either hypothesis $h \in \{0,1\}$,
\begin{align*}
\Prp{ T_U \leq \Delta_k \mid H_h^{(N_k)}} & = \Prp{ \bigcup_{t \leq \Delta_k} \{ U_t^{(N_k)} \geq b_t^{(N_k)} \} \mid H_h^{(N_k)} } \\
& = \Prp{ \max_{t \leq \Delta_k} \{ U_t^{(N_k)} - b_t^{(N_k)} \} \geq 0  \mid H_h^{(N_k)}}.
\end{align*}
Hence by \eqref{eq:proof:cor:impossible} and considering the maximal elements, we may find indices $\{k_m\}_{m=1,2,\ldots}$ and $t_m \leq \Delta_{k_m}$ such that
\begin{align}
    \label{eq:proof:cor:impossible:1}
\Prp{  U_{t_m}^{(N_{k_m})} \geq b_{t_m}^{(N_{k_m})}  \mid H_1^{(N_{k_m})}} - \Prp{ U_{t_m}^{(N_{k_m})} \geq b_{t_m}^{(N_{k_m})}  \mid H_0^{(N_{k_m})}} \geq a/2.
\end{align}
Strictly speaking, we construct $\{k_m\}_{m=1,2,...}$ by first finding a sub-sequence $\{k_{s}\}_{s=1,2\ldots}$ along which, say, 
\begin{align*}
\Prp{ \max_{t \leq \Delta_{k_s}} \{ U_t^{(N_{k_s})} - b_t^{(N_{k_s})} \} \geq 0 \mid H_1^{(N_{k_s})}} - 
\Prp{  U_{t_s}^{(N_{k_s})} \geq b_{t_s}^{(N_{k_s})} \mid H_0^{(N_{k_s})} } \geq  3a/4,
\end{align*}
and then find a subsequence of $\{k_{s}\}_{s=1,2\ldots}$ under $H_1^{(N_{k_s})}$ in a similar fashion. 

By \eqref{eq:dominance}, and since the distribution of $\tilde{U}$ and any $U_t$ are identical under the null, \eqref{eq:proof:cor:impossible:1} implies
\begin{align*}
    \Prp{ \tilde{U}^{(N_{k_m})} \geq b_{t_m}^{(N_{k_m})}  \mid H_1^{(N_{k_m})}} - \Prp{ \tilde{U}^{(N_{k_m})} \geq b_{t_m}^{(N_{k_m})} \mid H_0^{(N_{k_m})}} \geq a/2.
\end{align*}
Setting $b^{(N_{k_m})} = b_{t_m}^{(N_{k_m})}$, we obtain a contradiction to \eqref{thm:Kipnis2021_impossible}. 
\end{proof}

\begin{proof}[Proof of Corollary \ref{cor:possible_general}]
For every $t$ such that $\rho(\Delta) \geq \rho^+$, the array of distributions $\{Q_{n,t},\,n=1,\ldots,N,\, t \in \Theta_1\}$ 
is asymptotically stochastically bounded from above by the sequence of distributions $\{\tilde{Q}_n^{(N)}\, n=1,\ldots,N\}$ that satisfies \eqref{eq:logchisqaured_appx} with $\rho = \rho^+$. Denote by $\tilde{\pi}_1,\ldots, \tilde{\pi}_N$ P-values obeying \eqref{eq:hyp_test_pvalues} with $Q_{n}^{(N)} = \tilde{Q}_n^{(N)}$ and by $\tilde{\HC}^*$ the higher criticism of $\tilde{\pi}_1,\ldots,\tilde{\pi}_{N}$. Note that $\HC_t^*$ stochastically dominates $\tilde{\HC}^*$ under $H_1^{(N)}$ if $\rho(\Delta) \geq \rho^+$ and stochastically identical to $\tilde{\HC}^*$ under $H_0^{(N)}$ . By \eqref{thm:Kipnis2021_possible}, since 
$\rho^+ > \rho^*(\beta, \sigma)$, there exists $\{\tilde{b}^{(N)}\}$ such that 
\begin{align}
    \label{eq:cor_possible_H0}
\Prp{ \tilde{\HC}^* \leq \tilde{b}^{(N)} \mid H_1^{(N)} } + \Prp{ \tilde{\HC}^* > \tilde{b}^{(N)} \mid H_0^{(N)} } \to 0.
\end{align}
Set $b_t^{(N)} = \tilde{b}^{(N)}$ (fixed for all $t$) and consider the detection procedure $T_{\HC} = \inf\{ t\,:\, \HC_t^* \geq \tilde{b}^{(N)}\}$. 
Suppose by contradiction that 
\begin{align}
\limsup_{N \to \infty} \sup_{t \in \Theta_1} \Prp{T_{\HC} \leq t \mid  H_0^{(N)}} = a,
\label{eq:proof_cor_possible_limsup}
\end{align}
for some $a \geq 0$. This implies that there exists a sub-sequence $N_1<N_2<\ldots\leq N_k$, $k=1,2,\ldots$ such that 
\[
\sup_{t \in \Theta_1} \Prp{ \inf_{s\leq t} \{\HC^*_s\} > \tilde{b}^{(N_k)} \mid H_0^{(N_k)} } \geq 3a/4,
\]
and thus a sequence of times $t_1,t_2,\ldots$ (not necessarily in $\Theta_1$) such that 
$\Prp{ \HC^*_{t_k} > b^{(N_k)} \mid H_0^{(N_k)} } \geq a/2$. However, since each $\HC^*_{t_k}$ has distribution identical to that of $\tilde{\HC}^*$ under $H_0^{(N)}$, the previous assertion contradicts \eqref{eq:cor_possible_H0} if $a>0$. It follows that $a=0$ in  \eqref{eq:proof_cor_possible_limsup} thus $\Prp{T_{\HC} \leq \Tcal \mid H_0^{(N)}} \to 0$. Finally, for any $t$ such that $\rho(\Delta) > \rho^*(\beta, \sigma)$, 
\begin{align*}
    \Prp{ T_{\HC} \geq t \mid H_1^{(N)}} & \leq 
    \Prp{ \HC^*_t \leq b_t^{(N)} \mid H_1^{(N)}} =  
    \Prp{ \HC^*_t \leq \tilde{b}^{(N)} \mid H_1^{(N)}} \\
    & \leq \Prp{ \tilde{\HC}^* \leq \tilde{b}^{(N)} \mid H_1^{(N)}} \to 0. 
\end{align*}
The result follows because the convergence of either term is uniform in $t$. 
\end{proof}

\begin{proof}[Proof of Theorem~\ref{thm:alternative}]

Consider the case $\square = \LR$. Note that
\begin{align*}
V_{t,\tau} & \overset{D}{=} \left(\sqrt{\Delta}\sigma Z + \frac{\Delta}{2}\mu_N(r)\right)\mu_N(r) = \frac{1}{2}\left(\sigma Z + \sqrt{\Delta}\mu_N(r) \right)^2 - \sigma^2 Z^2/2  \\
& = \frac{1}{2}\left(\sigma Z + \mu_N(r\Delta) \right)^2 - \sigma^2 Z^2/2 \\
& = \frac{1}{2}\left(\sigma Z + \mu_N(r\Delta) \right)^2(1+o_p(1)),
\end{align*}
where in the last transition we used that $\mu_N(r) \to \infty$ as $N\to \infty$ and that 
\[
\frac{Z^2}{\mu_N^2(r(\Delta))} \sim N^{-r \Delta} \to 0.
\]
In particular, $V_{t,\tau}$ is unbounded in $N$. By Lemma~\ref{lem:argmax_GLR}, 
\[
Y_t^{\LR} = V_{t,\tau} + o_p(1), 
\]
From the above and using Lemma~\ref{lem:null_LR}, 
\begin{align*}
    -2\log \pi_t^{\LR}(Y_t^{\LR}) &= -2Y_t^{\LR}(1+o_p(1)) \\
    & = -2(V_{t,\tau} + o_p(1))(1+o_p(1)) = -2(V_{t,\tau})(1+o_p(1)) \\
    & = \left(\sigma Z + \mu_N(r\Delta) \right)^2 (1 + o_p(1)). 
\end{align*}

Next, consider the case $\square = \GLR$. We have
\[
W_{t,\tau}^2 \overset{D}{=} \left(\sqrt{\Delta}\mu_N(r) + \sigma Z  \right)^2 = \left(\mu_N(r\Delta) + \sigma Z  \right)^2,\qquad Z\sim \Ncal(0,1). 
\]
By Lemma~\ref{lem:argmax_GLR}, 
\[
(Y_t^{\GLR})^2 = W_{t,\tau}^2 + o_p(1) \overset{D}{=} \left(\mu_N(r\Delta) + \sigma Z \right)^2 +  o_p(1), 
\]
where $o_p(1)$ represents a RV that goes to $0$ in probability as $N \to \infty$, uniformly in $t$. From here, Lemma~\ref{lem:null_GLR}, and continuous mapping, 
\begin{align*}
-2\log \pi_t^{\GLR}(Y_t^{\GLR}) & \overset{D}{=} (Y_t^{\GLR})^2(1+o_p(1))  \\
& \overset{D}{=} \left(\mu_N(r\Delta) + \sigma Z  \right)^2(1 + o_p(1)). 
\end{align*}
In the last transition, we used $\mu_N(r) \to \infty$ as $n \to \infty$. 
\end{proof}

\section{Properties of change-point detection statistics under the null}
The results in this section were not used to prove our main results. They are provided here for readers interested in analytic expressions for survival functions of $Y_t^{\square}$, $\square \in \{\LR, \GLR\}$ that can serve as an alternative to obtaining the distribution of $Y_t^{\square}$ by Monte Carlo simulations involving multiple draws of an individual sequence $\{X_t\}$. 

A considerable effort in the literature on change point detection focused on characterizing the ``steady state'' null behavior of $Y_t^{\square}$. The following two lemmas use such results to provide the asymptotic behavior of $\pi_t^{\square}$ as $t \to \infty$, from which we may obtain asymptotic P-values as an alternative to simulations. These results show that  
\begin{align*}
\tilde{\pi}_t^{\LR}(Y_{t}^{\LR}) = e^{-Y_t^{\LR}} \quad \text{ and } \quad 
\tilde{\pi}_t^{\GLR}(Y_{t}^{\GLR})  = e^{-(Y_t^{\GLR})^2/2}, 
\end{align*} 
are asymptotic P-values under \eqref{eq:data_model}. 
The results on the asymptotic power of HC from \cite{kipnis2021logchisquared} apply to these asymptotic P-values.  
\begin{lemma}
\label{lem:null_LR}
    Consider $\pi_t^{\LR}$ of \eqref{eq:p-val_def_LR}. As $x \to \infty$ 
    and
    $\log(t)/x \to 0$,
    \begin{align}
\label{eq:null_LR}
             -2\log\pi_t^{\LR}(x) 
             = 2x(1+o(1))
\end{align}
uniformly in $t$. 
\end{lemma}

\begin{lemma}[Theorem 1 in \cite{siegmund1995using}] As $x\rightarrow \infty$, under $H_0$, 
$T_x^{\GLR}$ is asymptotically exponentially distributed with expectation
\begin{align}
    \label{eq:exp_Tb}
\ex{T_x^{\GLR}} \sim \frac{(2\pi)^{1/2} \exp(x^2/2)}{x\int_0^\infty z \nu^2(z) dz},
\end{align}
where the special function 
\[
\nu(x) = \frac{2}{x^2} \exp \left\{
-2\sum_{n=1}^\infty \frac 1n \Phi\left(-\frac{x\sqrt n}{2}\right)
\right\},
\]
with $\Phi$ the standard normal cumulative distribution function.
\label{thm:Siegmond1995}
\end{lemma}

\begin{lemma}
\label{lem:null_GLR}
Suppose that $x \to \infty$ and $t\to \infty$ such that $t \leq c x^2$ for some $c>0$. Then under $H_0$,
\begin{align}
    \label{eq:null_GLR}
-2\log \pi_t^{\GLR}(x) = x^2(1 + o(1)),
\end{align}
where $o(1) \to 0$. 
\end{lemma}

\begin{proof}[Proof of Lemma~\ref{lem:null_LR}]
A standard result in change-point detection says that as $x \to \infty$ and $t\to \infty$ with $\log(t)/x \to 0$, 
\begin{align*}
e^{x}\Prp{Y_t^{\LR} \geq x}/t \to C
\end{align*}
for some constant $C$. 
(c.f. \cite[Ch 4.3]{yakir2013extremes}). Consequently,
\begin{align*}
    -2\log \pi_t^{\LR}(x) & = -2\log \Prp{Y_t^{\LR} \geq x} \\
    & = 2x - 2\log(Ct)+o(1) \\
    & = 2x(1 - \log(C)/x - \log(t)/x) + o(1) \\
    & = 2x(1+o(1))
\end{align*}
\end{proof}

\begin{proof}[Proof of Lemma \ref{lem:null_GLR}]
It follows from Lemma~\ref{thm:Siegmond1995} that under $H_0$ and $t/x^2 = O(1)$, $T_x$ is asymptotically exponentially distributed with mean
\[
\mu_x := \ex{T_x|H_0} = \sqrt{2\pi}\frac{e^{x^2/2}}{x C},
\]
where $C$ is some numerical constant. Therefore, for $x>0$ we may write 
\begin{align*}
\Prp{Y_t^{\GLR} \geq x} & = 
\Prp{ T_x \leq t} \\
& =
(1 - e^{-t/\mu_{x}})(1 + o(1)) = t/\mu_x(1+o(1)).
\end{align*}
It follows that
\begin{align*}
    -2\log\Prp{Y_t^{\GLR} \geq x} 
    & =
    2 \log(\mu_x) - 2 \log(t) + o(1) \\
    & = x^2 + 2\log \frac{\sqrt{2\pi}}{xC} - 2\log(t) + o(1) \\
    & = x^2(1+o(1)).
\end{align*}
\end{proof}

\end{appendix}

\end{document}